\documentclass[]{aastex631}

\usepackage{epstopdf}

\begin{document}

\title{Asteroseismology and Dynamics Reveal Interior Structure and Coeval Evolution in the Triply Post-Main-Sequence system DG Leo}

\correspondingauthor{Wen-Ping Liao}
\email{$^{*}$liaowp@ynao.ac.cn}	
\author{Ping Li}
\affiliation{Yunnan Observatories, Chinese Academy of Sciences (CAS), 650216 Kunming, China}
\affiliation{University of Chinese Academy of Sciences, No.1 Yanqihu East Rd, Huairou District, Beijing, China 101408}

\author{Wen-Ping Liao$^{*}$}
\affiliation{Yunnan Observatories, Chinese Academy of Sciences (CAS), 650216 Kunming, China}
\affiliation{University of Chinese Academy of Sciences, No.1 Yanqihu East Rd, Huairou District, Beijing, China 101408}

\author{Sheng-Bang Qian}
\affiliation{School of Physics and Astronomy, Yunnan University, Kunming 650091, China}

\author{Li-Ying Zhu}
\affiliation{Yunnan Observatories, Chinese Academy of Sciences (CAS), 650216 Kunming, China}
\affiliation{University of Chinese Academy of Sciences, No.1 Yanqihu East Rd, Huairou District, Beijing, China 101408}

\author{Jia Zhang}
\affiliation{Yunnan Observatories, Chinese Academy of Sciences (CAS), 650216 Kunming, China}

\author{Qi-Bin Sun}
\affiliation{School of Physics and Astronomy, Yunnan University, Kunming 650091, China}

\author{Fang-Bin Meng}
\affiliation{Yunnan Observatories, Chinese Academy of Sciences (CAS), 650216 Kunming, China}
\affiliation{University of Chinese Academy of Sciences, No.1 Yanqihu East Rd, Huairou District, Beijing, China 101408}



\begin{abstract}

$\delta$ Scuti stars in binary or multiple systems serve as crucial probes for studying stellar pulsation and evolution. However, many such systems are not ideal for asteroseismology due to uncertainties in mass transfer with close companions and the challenges of dynamically measuring all components' physical properties. The triple system DG~Leo, comprising an inner binary and a distant  $\delta$ Scuti star, is an ideal target due to its well-separated pulsator. By combining new \textit{TESS} photometry with archival spectroscopy, our dynamical analysis shows that the system's three components share similar masses, radii, and luminosities within errors, occupying coincident Hertzsprung--Russell diagram positions, indicative of coeval evolution. By fitting seven observed $\delta$ Scuti frequencies through asteroseismic modeling with dynamically constrained theoretical grids, we simultaneously trace the pulsating star's evolution and constrain the triple system's evolutionary stage, with the derived fundamental parameters showing consistency with the dynamical solutions. Our analysis reveals that all three components of DG~Leo are in the post-main-sequence phase, with a system age of $0.7664^{+0.1402}_{-0.1258}$~Gyr. Additionally, the $\delta$ Scuti component shows multiple non-radial modes with significant mixed-character frequencies, providing precise constraints on its convective core extent ($R_{\mathrm{cz}}/R = 0.0562^{+0.0137}_{-0.0021}$). 

\end{abstract}



\section{Introduction} \label{sec:intro}
Systems hosting pulsating stars in binary or multiple configurations provide crucial laboratories for probing stellar deep structure and evolution. Recently, space missions, notably \textit{Kepler} \citep{2010Sci...327..977B} and \textit{TESS} \citep{2015JATIS...1a4003R}, supply high-precision, long-baseline photometry for such systems. Moreover, the spectroscopic data from facilities like \textit{LAMOST} enable determinations of component parameters including mass ratio, effective temperature, chemical composition, and surface gravity. Consequently, fundamental stellar parameters (e.g., mass, radius, luminosity) can be dynamically measured with sufficient accuracy to directly constrain stellar evolution models for pulsators.

 \textbf{$\delta$ Scuti stars} are A/F-type pulsators located at the intersection of the main sequence and the Cepheid instability strip in the Hertzsprung-Russell (H-R) diagram. Their oscillation periods typically range from 0.02 to 0.25~days \citep{2000ASPC..210....3B} and are primarily driven by the $\kappa$ mechanism operating in the helium second ionization zone \citep{1963ARA&A...1..367Z,1965ApJ...142..868B,1994A&A...286..815L,1971A&A....14...24C}. These stars exhibit both radial and non-radial pulsation modes, making them prime targets for asteroseismic investigations. Furthermore, a subset known as high-amplitude $\delta$ Scuti stars (HADS) displays fewer radial modes with high amplitudes, manifesting as single-periodic variations in the radial fundamental or first overtone, as well as double- or triple-periodic pulsations \citep{2022MNRAS.515.4574N}. Many binary or multiple systems hosting $\delta$ Scuti stars are presented \citep{2012MNRAS.422.1250L, 2018MNRAS.475..478Q, 2022ApJS..259...50S, 2022ApJS..263...34C}. For example, systems such as HZ Dra \citep{2024ApJ...977....3L}, HL Dra \citep{2021MNRAS.505.6166S} and CZ Aqr \citep{2024ApJ...975..249Z} have been reported to have a tertiary star around the inner close binary system. However, many of them are not ideal objects for probing stellar evolution using asteroseismology due to:  
i) uncertain mass transfer between the $\delta$ Scuti component and its close companions \citep{2020ApJ...895..136C};  
ii) observational challenges in dynamically determining all components' physical properties via combined spectroscopy, photometry, or astrometry. 

DG~Leo (TIC~88024537) is a hierarchical and spectroscopic triple system consisting of a inner binary (components Aa and Ab) and a distant companion (component B) forming a wider visual binary. The close binary components both exhibit spectral types A8 IV \citep{1982bsc..book.....H}, while the composite spectrum has been classified as F0 IIIn \citep{1976PASP...88...95C} and A7 III \citep{1979PASP...91...83C}. The orbital period of the Aa,b system is 4.147 days \citep{1977PASP...89..216F}, while the orbital period of the visual pair A \& B is approximately 200 years. The visual component B, observed by \textit{Hipparcos} with an angular separation of $0''.17$ and a magnitude difference of 0.69 mag relative to Aa,b \citep{1997ESASP1200.....E}, was monitored through micrometric observations and speckle interferometry from 1935 to 1997 \citep{2000AJ....119.3084H}. Existing astrometric data indicate a highly inclined and potentially eccentric orbit. Previous studies \citep{1977PASP...89..658B,1991PASP..103..628R} presented evidence for possible shallow eclipses with an inclination of $\sim$75$^{\circ}$ for the Aa,b system. 

All three components of this system reside within the $\delta$ Scuti instability strip, rendering them potential pulsators. \cite{1967ApJ...149...55D} first reported short-period variability in the system but could not determine its nature. Furthermore, claims regarding short-period oscillations primarily concern component B, which is classified as both an ultra-short-period Cepheid \citep{1979ApJS...41..413E} and a $\delta$ Scuti star \citep{1974AJ.....79.1082E}, with possible changes in amplitude and phase reported by \cite{1982IBVS.2152....1A,1991MNRAS.252...68R}. Two decades ago, the spectroscopic analysis resolved this ambiguity: the inner binary consists of two Am stars, at least one rotating asynchronously despite a circular orbit, whereas component B is a chemically normal $\delta$ Scuti pulsator \citep{2005MNRAS.356..545F}. Ellipsoidal variations from tidally deformed components were first detected by \cite{2004RMxAC..21...73L,2005A&A...438..201L}.

In summary, despite extensive studies on DG~Leo, key physical properties such as masses of its three component stars remain undetermined dynamically. Recently, the \textit{TESS} mission observed DG~Leo across four sectors (Sectors~21, 45, 46, and~72), providing continuous 120-second cadence light curves (All of the \textit{TESS} data used in
this paper can be found through the MAST: \dataset[doi:10.17909/k0fp-8v04]{https://doi.org/10.17909/k0fp-8v04}). When combined with the spectroscopic results of previous studies \citep{2004RMxAC..21...73L,2005A&A...438..201L}, this provides a unique opportunity to determine the physical parameters (e.g., mass, radius) of all three components dynamically. It is then possible to infer the evolution of the triple system by investigating the component B using asteroseismology. This also enables us to probe the interior structure of component B. Thanks to its clear separation and these sufficient observations, DG~Leo is therefore a good target for studying stellar evolution using asteroseismology.

The paper is structured as follows. Section \ref{sec:WD} details the modeling of the \textit{TESS} light curve and derivation of stellar parameters. Section \ref{sec:fre} presents the analysis of pulsational characteristics in DG~Leo. In Section \ref{sec:stellar models}, we describe the construction and analysis of the stellar model for DG~Leo. Finally, Section \ref{sec:dc} provides the main discussions and conclusions.

\section{Modeling the inner binary system} \label{sec:WD}
\begin{figure}[ht!]
	\plotone{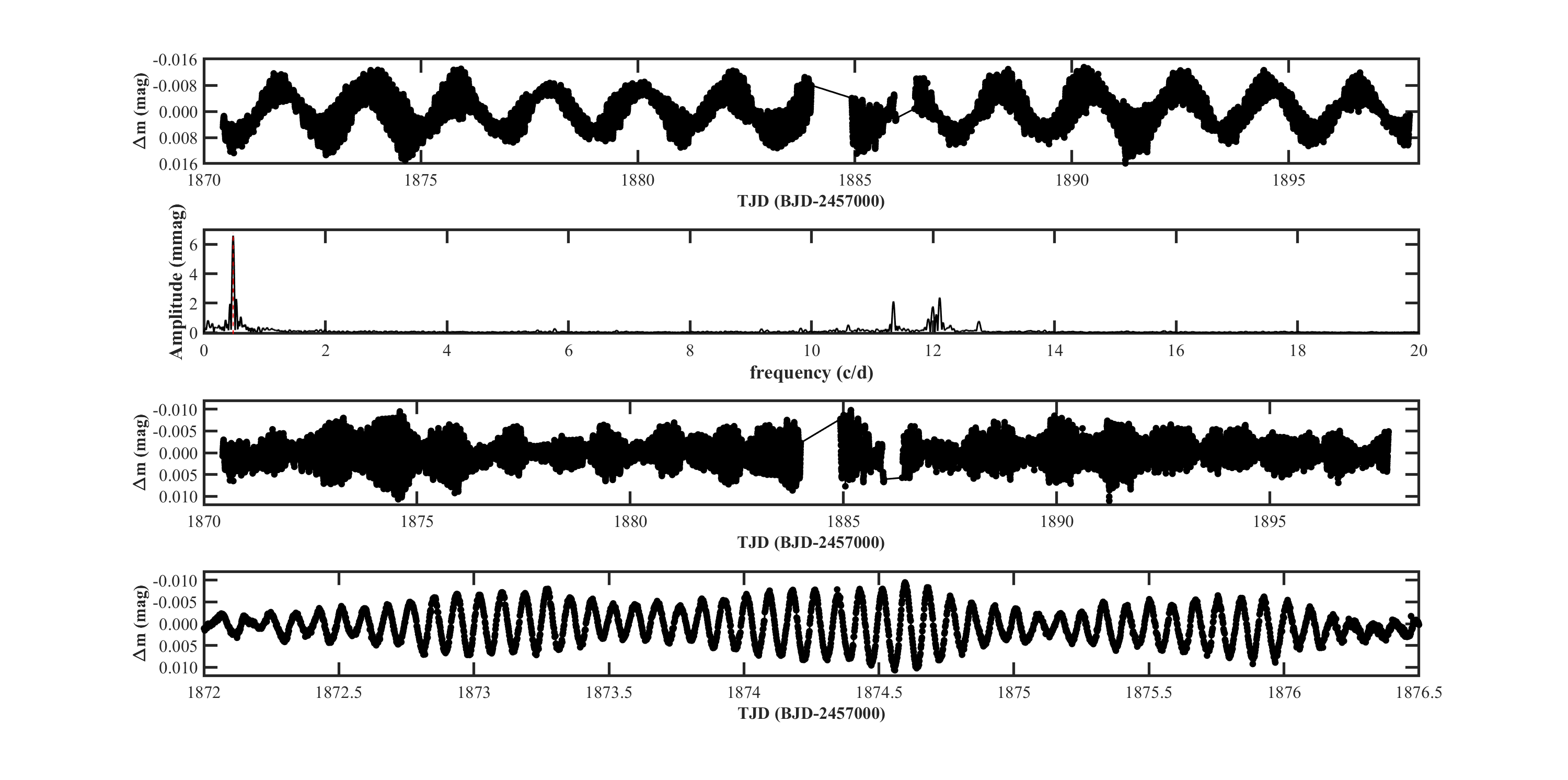}
	\caption{The \textit{TESS} light curve and corresponding Fourier amplitude spectrum of DG~Leo. Top panel: the full light curve from Sector 21. Second panel: the Fourier amplitude spectrum derived from the full light curve. The dominant frequency peak at 0.4822\,d$^{-1}$ (red dashed vertical line) corresponds to ellipsoidal variations, while other significant peaks represent pulsation frequencies. Third panel: the light curve after removing the dominant ellipsoidal variations. Bottom panel: a zoomed-in view of the detrended light curve between TJD 1872--1876.5. (TJD = BJD $-$ 2457000.0).}
		\label{fig:lc}
\end{figure}

As shown in the top and second panels of Figure \ref{fig:lc}, the ellipsoidal variations reported by \citet{2004RMxAC..21...73L,2005A&A...438..201L} are also present in the \textit{TESS} light curve. The dominant period of 2.0737\,d corresponds to half the orbital period of the inner spectroscopic binary. However, the signal with period of 4.166 d \citep{2004RMxAC..21...73L} is not determined in the \textit{TESS} light curve. Moreover, as shown in the bottom panel of Figure \ref{fig:lc}, the peak pulsation amplitude is about 0.01 mag. This is an order of magnitude smaller than the typical amplitude of HADS stars, which are defined by V-band variations exceeding 0.1 mag \citep{2002ApJ...576..963T}.

To derive reliable orbital parameters for this inner binary system, we modeled the light curves of four sectors using the Wilson--Devinney (W-D) method \citep{1971ApJ...166..605W,1979ApJ...234.1054W,1990ApJ...356..613W,2007ApJ...661.1129V,2012AJ....144...73W}. To minimize the influence of pulsation effects and enhance the data reliability, we generated an averaged light curve through a two-step process: (i) converting TJD values to phase units, and (ii) reducing the original dataset from over 18,998 data points to 200 representative points by averaging within each sector. Specifically, all points falling within a phase interval of 0.005 were combined into a single averaged point, thereby smoothing the curve while preserving key features.
\begin{figure}[ht!]
	\plotone{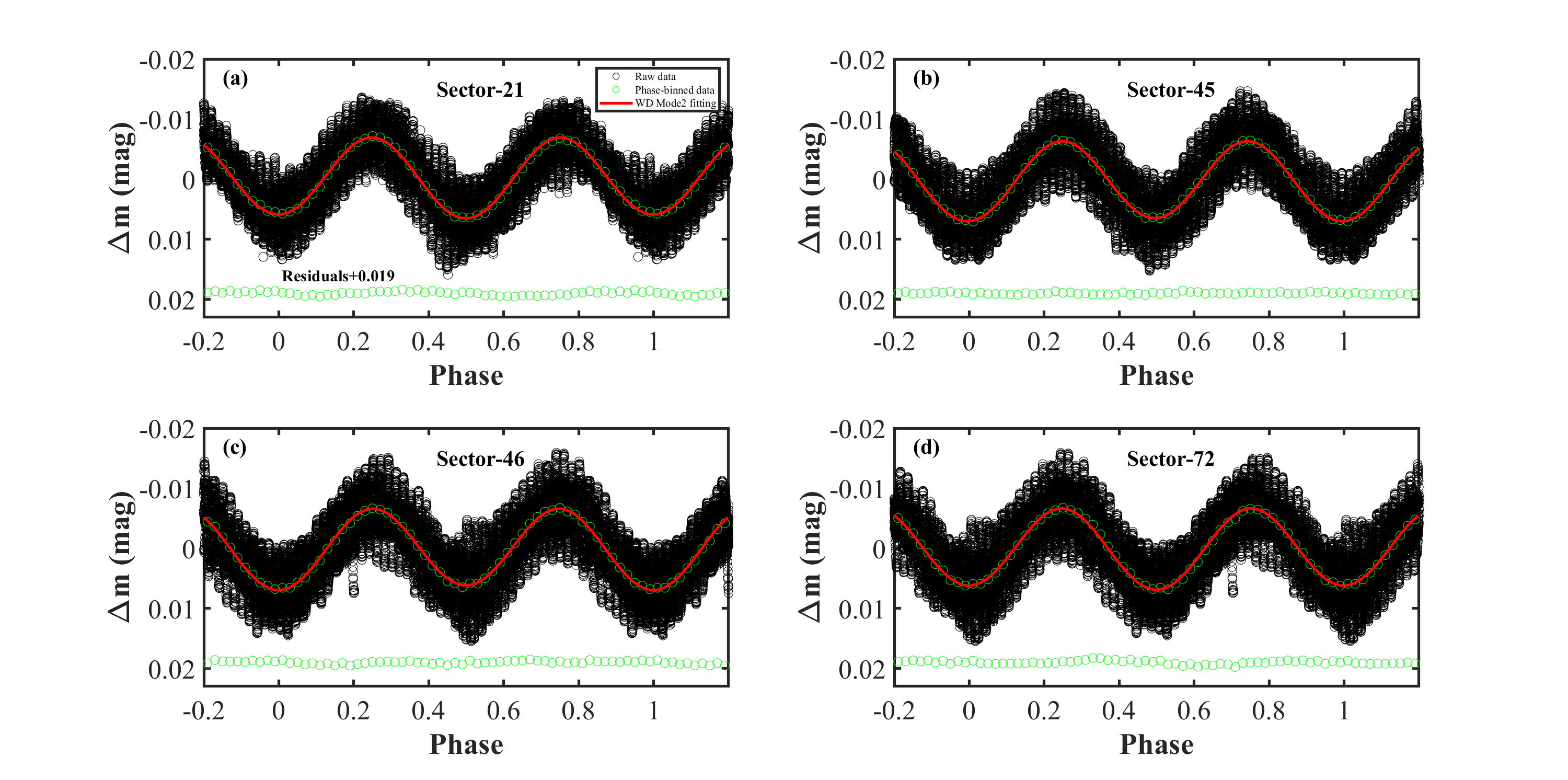}
	\caption{The light curves from four \textit{TESS} sectors and their corresponding phase-binned light curves (green circles) are overplotted with the best-fitting model (red solid curves; W-D Model~2). Residuals are shown in the bottom panel for each sector.
		\label{fig:wd}}
\end{figure}

Prior to light curve modeling, important parameters were fixed in the W-D code, with the primary star's effective temperature set to $7470 \pm 220$~K \citep{2005MNRAS.356..545F,2004ASPC..318..342F}. As both components are early-type stars, their bolometric albedos $A_1$ and $A_2$ were fixed at $1.00$ \citep{2001MNRAS.327..989C} and gravity-darkening coefficients $g_1$ and $g_2$ at $1.00$ \citep{1967ZA.....65...89L}, while the mass ratio $q = M_2/M_1$ was fixed at the spectroscopic value $1.000 \pm 0.001$ \citep{2005MNRAS.356..545F,2004ASPC..318..342F}. Linear limb-darkening coefficients $x_1$ and $x_2$ were adopted from \citet{1993AJ....106.2096V}. For the inner binary system, W-D Model~2 (detached configuration) was implemented, with orbital eccentricity and period fixed at $e = 0$ and $P = 4.146751$~d respectively \citep{2005MNRAS.356..545F}, while free parameters—including primary luminosity ($L_1$), orbital inclination ($i$), primary potential ($\Omega_1$), secondary potential ($\Omega_2$), and secondary temperature ($T_2$)—were iteratively adjusted during optimization.
\begin{table*}
	\scriptsize
	\begin{center}
		\caption{Light curve modeling solutions for four sector light curves of DG~Leo. The numbers in brackets are the errors in the last two bits of the data. The units of most parameters are dimensionless, except for those explicitly specified.\label{tab:solutions}}
		\begin{tabular}{lccccc}
			\hline
			Parameters & Sector 21 &  Sector 45 &  Sector 46 &  Sector 72& Mean values \\
			\hline
			Period (d) & 4.146751 & 4.146751 & 4.146751 & 4.146751 & 4.146751 \\
			$i$ ($^{\circ}$) & 66.55(95) & 66.0(13) & 66.0(18) & 68.91(48) & 66.87(60) \\
			$q$ ($M_2/M_1$) & 1.000 & 1.000 & 1.000 & 1.000 &1.000 \\
			$T_1$ (K) & 7470 & 7470 & 7470 & 7470 & 7470 \\
			$T_2/T_1$ & 0.9705(53) & 0.9683(55) & 1.024(41) & 1.030(48) & 0.998(17) \\
			$L_1/(L_1+L_2)_{\rm TESS}$ & 0.5286(46) & 0.5242(39) & 0.486(31) & 0.523(33) & 0.515(10) \\
			$L_2/(L_1+L_2)_{\rm TESS}$ & 0.4714(46) & 0.4758(39) & 0.514(31) & 0.477(33) & 0.485(10) \\
			$\Omega_1$ & 6.192(38) & 6.415(51) & 6.41(17) & 6.29(12) & 6.327(53) \\
			$\Omega_2$ & 5.28(41) & 6.434(40) & 6.43(20) & 6.76(33) & 6.226(325) \\
			$r_1$(pole)$^b$ & 0.1834(28) & 0.1841(17) & 0.1841(58) & 0.1884(41) & 0.185(1)\\
			$r_1$(point)$^b$ & 0.1859(32) & 0.1874(18) & 0.1874(62) & 0.1920(44) & 0.188(1)\\
			$r_1$(side)$^b$ & 0.1844(29) & 0.1853(18) & 0.1853(59) & 0.1897(42) & 0.186(1)\\
			$r_1$(back)$^b$ & 0.1855(31) & 0.1869(18) & 0.1869(61) & 0.1915(44) & 0.188(1)\\
			$r_2$(pole)$^b$ & 0.180(23) & 0.1834(33) & 0.1835(66) & 0.1733(98) & 0.180(2) \\
			$r_2$(point)$^b$ & 0.184(25) & 0.1866(75) & 0.1867(71) & 0.176(10) & 0.183(3)\\
			$r_2$(side)$^b$ & 0.181(23) & 0.1846(30) & 0.1846(68) & 0.174(10) & 0.181(2)\\
			$r_2$(back)$^b$ & 0.183(25) & 0.1862(45) & 0.1862(70) & 0.175(10) & 0.183(3)\\
			$f_1$ (Filling degree of the primary) & 0.0952(75) & 0.1173(18) & 0.1173(59) & 0.1259(44) & 0.114(7)\\
			$f_2$ (Filling degree of the second) & 0.134(29) & 0.1160(21) & 0.1160(67) & 0.0974(89)& 0.116(7) \\
			\hline
			\hline
			\multicolumn{2}{l}{\textbf{Absolute parameters}} \\  % 合并两列，加粗标题
			\hline
			\hline
			$M_1$ (M$_{\odot}$) &2.258(33)&&$M_B$ (M$_{\odot}$) &2.39(40)\\
			$M_2$ (M$_{\odot}$) &2.258(33)&&......\\
			$R_1$ (R$_{\odot}$) &3.353(15)&&$R_B$(R$_{\odot}$) & 2.95(50)\\
			$R_2$ (R$_{\odot}$) &3.263(25)&&......\\
			$L_1$ (L$_{\odot}$) &25.99(1.44)&&$L_B$ (L$_{\odot}$) &26.05(8.42)\\
			$L_2$ (L$_{\odot}$) &24.48(1.67) &&......\\
			$a$ (R$_{\odot}$) &17.955(62)&&......\\
			\hline
		\end{tabular}
	\end{center}
	\footnotesize $^{b}$ In units of semi-major axis.\\
\end{table*}

These free parameters were iteratively adjusted until achieving converged solutions. The results for all four sectors' light curves are listed in the second, third, fourth, and fifth columns of Table \ref{tab:solutions}. Based on these solutions, theoretical light curves were calculated as depicted by the red curves in Figure \ref{fig:wd}. The light-curve residuals for each solution are plotted in the bottom panel for each sector (green circles). All four converged solutions reveal that the inner binary of DG~Leo is a detached system consisting of two A-type components with low Roche lobe filling factors.

Combining the spectroscopic solution \citep{2004RMxAC..21...73L,2005MNRAS.356..545F} with the mean values of the light-curve modeling results, we dynamically determined the physical parameters (e.g, mass, radius, luminosity, and separation) of both components as listed in Table \ref{tab:solutions}. Based on the mass ratio of 0.374$\pm$0.046 between the distant component B and the total system mass—determined by \cite{2005MNRAS.356..545F}, we calculated the mass of component B and included it in Table \ref{tab:solutions}. Additionally, we estimated the radius and luminosity of component B, with the radius agreeing well with $R_B = 2.99\pm0.20$ R$_{\odot}$ from \citet{2005MNRAS.356..545F}.

\section{Frequency Analysis} \label{sec:fre}
\textit{TESS} observed DG Leo continuously from TJD 2526.0097 to TJD 2578.7048, generating data files s45–s46. To analyse the pulsation properties of DG Leo, we subtracted the light curve model from these \textit{TESS} light curves. The frequency resolution for this dataset is $\delta f = 1/\Delta T \approx 0.0189$ c/d, where $\Delta T$ is the total observational time span for sectors s45-s46. We conducted a detailed analysis for investigating the pulsation features through multiple frequency analysis of the light curve residuals using the \textit{Period04} software \citep{2005CoAst.146...53L}. Since no pulsation signals were detected in the high-frequency region ($f > 100$ c/d), subsequent frequency extraction was confined to the range 0–100 c/d.

\begin{figure}[ht!]
	\plotone{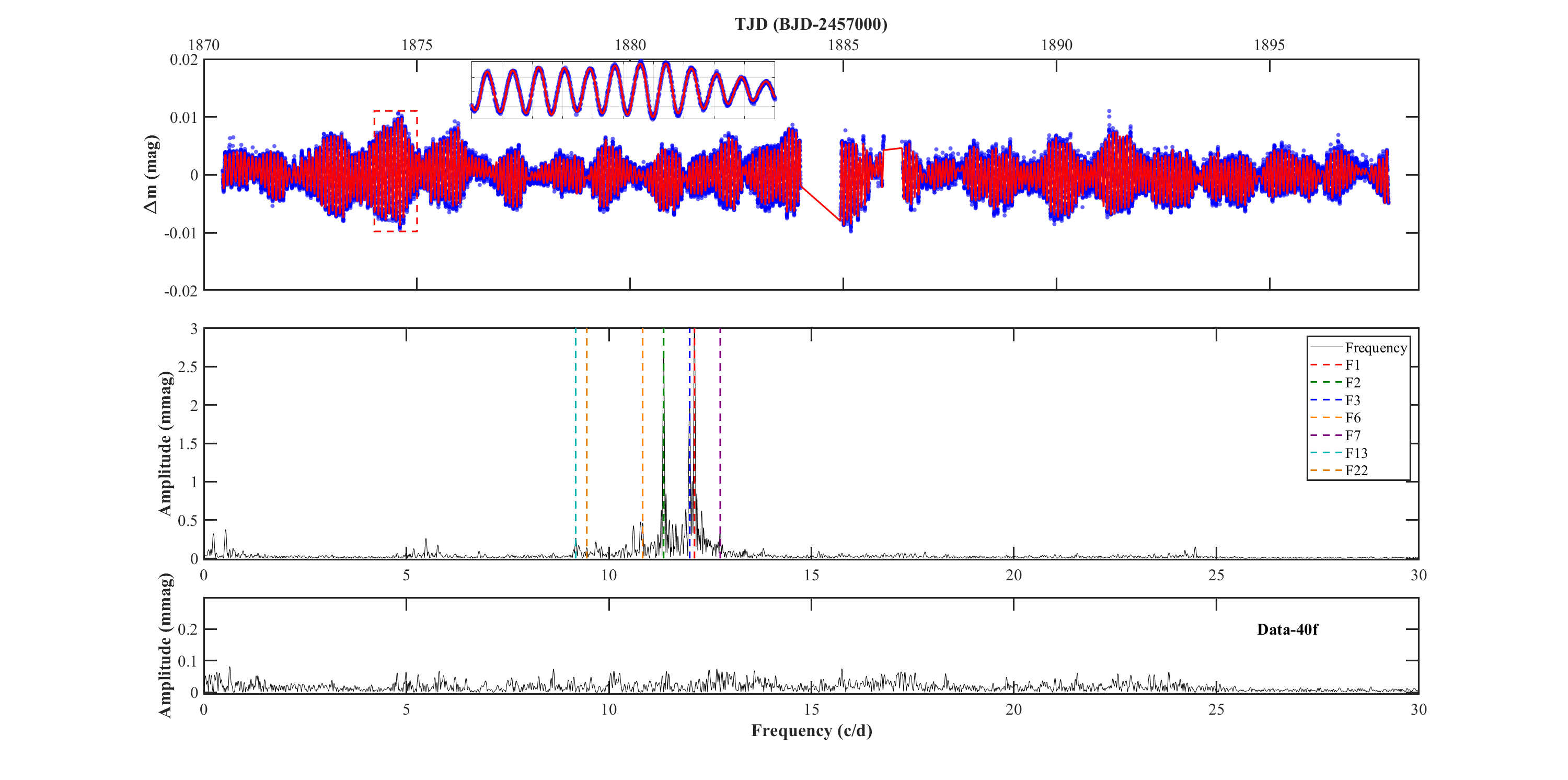}
	\caption{The light curve of DG Leo after subtracting the binary model, along with its corresponding Fourier spectra. Top panel: the residual light curve (blue dots) with the Fourier fit model overplotted (red curve). The inset shows a zoomed-in view of the same light curve between TJD 1872 and 1874. Middle panel: the Fourier amplitude spectrum of the residual light curve; the seven independent frequencies are marked by colored vertical dashed lines. Bottom panel: the residual spectrum after pre-whitening with 40 significant frequencies.} \label{fig:spec}
\end{figure}

In each iteration step, we identified the frequency with the highest amplitude and performed a multi-period least-squares fit to the data, incorporating all previously detected frequencies. The data were then pre-whitened using this solution, and the resulting light curve residuals served as the basis for further analysis. This process continued iteratively until the residuals met the signal-to-noise threshold of S/N $\geq$ 4.0 \citep{1993A&A...271..482B}. Ultimately, a total of 40 frequencies were detected, of which 38 satisfied the S/N $\geq$ 4.0 criterion; these are listed in Table \ref{tab:table 2}. Based on the method described by \cite{1999DSSN...13...28M}, we calculate the noise within a 1 c/d bandwidth centered on each frequency. We also determine the corresponding amplitudes and their uncertainties. Figure \ref{fig:spec} displays the light curve residuals after the removal of the binary model (blue dots) with the Fourier fit (red curve) and its corresponding Fourier spectrum. The top panel reveals pulsations with a maximum amplitude of 0.010 mag, which is substantially lower than the 0.016 mag amplitude of the ellipsoidal variation that was removed (see top panel of Figure~\ref{fig:lc}).

We examined the extracted frequencies to identify linear combination frequencies of the form $F_{k} = m F_{i} + n F_{j}$ \citep{2012AN....333.1053P,2015MNRAS.450.3015K}, where $m$ and $n$ are integers, $F_{i}$ and $F_{j}$ represent parent frequencies, $F_{k}$ denotes the combination frequency. A peak is classified as a combination if it satisfies two criteria: (1) both parent frequencies have amplitudes greater than the presumed combination term, and (2) the difference between observed and predicted frequencies is smaller than the frequency resolution $\delta f$ \citep{1978Ap&SS..56..285L,2019AJ....157...17L}. As summarized in Table \ref{tab:table 2}, we identified 24 linear combination frequencies, 6 possible side lobe, and one unknown frequency. Moreover, seven independent frequencies were identified: $F_1$, $F_2$, $F_3$, $F_6$, $F_7$, $F_{13}$, and $F_{22}$. These are indicated by the colored vertical dashed lines in Figure~\ref{fig:spec}. The seven independent frequencies range from 9 to 13 c/d (corresponding to periods of 0.07 to 0.1 d). This period range falls well within the typical period range known for $\delta$ Scuti stars (e.g., 0.02-0.25 d, \cite{2000ASPC..210....3B}).
\begin{deluxetable*}{lcccclcccc}
	\tabletypesize{\scriptsize}
	\label{tab:table 2}
	\tablecaption{Pulsating frequencies of DG Leo. The numbers in parentheses are the errors on the last bit of the data. The seven independent frequencies are indicated by bold formatting. \textbf{Frequencies marked as 'Combination' represent higher-order linear combinations of these independent frequencies.}}
	\tablehead{
		\colhead{ID} & \colhead{Freq. (c/d)} & \colhead{Ampl. (mag)} & \colhead{Remark} & \colhead{S/N} & 
		\colhead{ID} & \colhead{Freq. (c/d)} & \colhead{Ampl. (mag)} & \colhead{Remark} & \colhead{S/N}
	} 
	\startdata
	\textbf{\textit{F}\textsubscript{1}} & \textbf{12.112052 (61)} & \textbf{0.003057 (8)} &  & \textbf{113.96} & 
	\textit{F}\textsubscript{20} & 11.084639 (100) & 0.000186 (8) & Sidelobe & 9.05 \\
	\textbf{\textit{F}\textsubscript{2}} & \textbf{11.350499 (70)} & \textbf{0.002667 (8)} &  & \textbf{143.58} & 
	\textit{F}\textsubscript{21} & 9.820529 (108) & 0.000172 (8) & $3F_2-2F_1$ & 8.83 \\
	\textbf{\textit{F}\textsubscript{3}} & \textbf{11.993720 (92)} & \textbf{0.002017 (8)} &  & \textbf{79.08} & 
	\textbf{\textit{F}\textsubscript{22}} & \textbf{9.453882 (115)} & \textbf{0.000161 (8)} &  & \textbf{7.83} \\
	\textit{F}\textsubscript{4} & 11.925598 (354) & 0.000525 (8) & Sidelobe & 21.36 & 
	\textit{F}\textsubscript{23} & 24.475474 (122) & 0.000152 (8) & $4F_1-2F_3$ & 9.82 \\
	\textit{F}\textsubscript{5} & 12.282446 (359) & 0.000516 (8) & $2F_1-F_4$ & 19.57 & 
	\textit{F}\textsubscript{24} & 9.268517 (126) & 0.000147 (8) & \textbf{Combination} & 6.99 \\
	\textbf{\textit{F}\textsubscript{6}} & \textbf{10.833761 (432)} & \textbf{0.000430 (8)} &  & \textbf{19.81} & 
	\textit{F}\textsubscript{25} & 0.110737 (160) & 0.000116 (8) & $F_1-F_3$ & 4.92 \\
	\textbf{\textit{F}\textsubscript{7}} & \textbf{12.748753 (487)} & \textbf{0.000381 (8)} &  & \textbf{13.77} & 
	\textit{F}\textsubscript{26} & 13.823227 (153) & 0.000121 (8) & Sidelobe & 4.94 \\
	\textit{F}\textsubscript{8} & 0.545826 (58) & 0.000322 (8) & \textbf{Combination} & 12.39 & 
	\textit{F}\textsubscript{27} & 10.328425 (149) & 0.000125 (8) & \textbf{Combination} & 6.18 \\
	\textit{F}\textsubscript{9} & 11.723486 (50) & 0.000371 (8) & \textbf{Combination} & 16.97 & 
	\textit{F}\textsubscript{28} & 5.184296 (165) & 0.000113 (8) & \textbf{Combination} & 5.91 \\
	\textit{F}\textsubscript{10} & 10.604561 (52) & 0.000358 (8) & $2F_2-F_1$ & 17.12 & 
	\textit{F}\textsubscript{29} & 24.225893 (169) & 0.000110 (8) & $2F_1$ & 6.37 \\
	\textit{F}\textsubscript{11} & 10.779992 (47) & 0.000396 (8) & \textbf{Combination} & 18.50 & 
	\textit{F}\textsubscript{30} & 9.708836 (150) & 0.000124 (8) & Sidelobe & 6.15 \\
	\textit{F}\textsubscript{12} & 0.236882 (58) & 0.000319 (8) & \textbf{Combination} & 13.50 & 
	\textit{F}\textsubscript{31} & 0.734395 (10) & 0.001872 (8) & Sidelobe & 79.97 \\
	\textbf{\textit{F}\textsubscript{13}} & \textbf{9.180254 (70)} & \textbf{0.000264 (8)} &  & \textbf{13.09} & 
	\textit{F}\textsubscript{32} & 15.179588 (185) & 0.000100 (8) & \textbf{Combination} & 5.47 \\
	\textit{F}\textsubscript{14} & 5.483229 (70) & 0.000264 (8) & \textbf{Combination} & 12.58 & 
	\textit{F}\textsubscript{33} & 6.796465 (190) & 0.000098 (8) & $7F_2-6F_1$ & 5.71 \\
	\textit{F}\textsubscript{15} & 10.424492 (81) & 0.000229 (8) & \textbf{Combination} & 11.04 & 
	\textit{F}\textsubscript{34} & 0.963339 (187) & 0.000099 (8) & \textbf{Combination} & 4.60 \\
	\textit{F}\textsubscript{16} & 9.683210 (86) & 0.000215 (8) & \textbf{Combination} & 10.76 & 
	\textit{F}\textsubscript{35} & 0.735726 (100) & 0.001853 (8) & Sidelobe & 79.20 \\
	\textit{F}\textsubscript{17} & 5.776923 (95) & 0.000196 (8) & \textbf{Combination} & 10.07 & 
	\textit{F}\textsubscript{36} & 9.900610 (215) & 0.000086 (8) & \textbf{Combination} & 4.38 \\
	\textit{F}\textsubscript{18} & 11.277650 (96) & 0.000194 (8) & \textbf{Combination} & 10.46 & 
	\textit{F}\textsubscript{37} & 0.505719 (119) & 0.000155 (8) & unknown  & 5.99 \\
	\textit{F}\textsubscript{19} & 12.075402 (89) & 0.000208 (8) & \textbf{Combination} & 8.03 & 
	\textit{F}\textsubscript{38} & 0.639501 (184) & 0.000101 (8) & $F_7-F_1$ & 4.09 \\
	\tableline
	\enddata
\end{deluxetable*}

For identifying the pulsation modes of the independent frequencies, we calculated the pulsation constant $Q$ values for the 7 independent frequencies using the following basic relation for pulsating stars:
\begin{eqnarray}\label{equation (1)}
	Q=\mathit{P}_{\rm osc}\sqrt{\frac{\overline{\rho}}{\overline{\rho}_{\odot}} },
\end{eqnarray}
where $\mathit{P}_{\rm osc}$ is the pulsation period \textbf{and mean density} is $\overline{\rho}=0.094 (\pm0.011) \overline{\rho}_{\odot}$. The mean density of component B is estimated based on the parameters in Table \ref{tab:solutions}. The values of $Q$ for each frequency are used for comparison with the theoretical models \citep{1981ApJ...249..218F} for $M$=2.0 M$_{\odot}$ and 2.5 M$_{\odot}$. The probable $l$-degrees and the types of each independent frequency are identified and \textbf{listed in Table \ref{tab:3}.}

\begin{deluxetable*}{lcccc}  
	\tabletypesize{\scriptsize}  
	\tablecaption{The independent pulsating frequencies detected in DG~Leo.}\label{tab:3}
	\tablehead{  
		\colhead{ID} & \colhead{$F_{\rm obs}$ (c/d
			)} & \colhead{$Q\times 100$} & \colhead{$l$-degrees} & \colhead{Mode}  
	}  
	\startdata  
	$F_1$  & 12.112052 (61) &2.53  &0& 1H \\
	$F_2$  & 11.350499 (70) &2.70  &2& NRP1 \\
	$F_3$ & 11.993720 (92) & 2.55  &3&NRP1\\
	$F_6$ & 10.833761 (432) & 2.82 &1, 2 or 3&NRP1 \\
	$F_7$ & 12.748753 (487) &2.40 &3&NRP1 \\
	$F_{13}$ & 9.180254 (70) &3.33&2 or 3&f or g \\
	$F_{22}$ & 9.453882 (115) &3.24 &0&F \\ 
	\tableline  
	\enddata  
	\tablecomments{  
		The labels denote pulsation modes as follows: F (fundamental radial mode), 1H (first overtone radial mode), NRP1 (non-radial mode with one order), f (f-mode), and g (g-mode). 
	}  
\end{deluxetable*}

As shown in Table~\ref{tab:3}, the values of Q indicate F$_1$ is a first-overtone radial mode while F$_{22}$ is fundamental. The period ratio between F$_1$ and F$_{22}$ (0.780) matches the theoretical fundamental/first-overtone value (0.772), confirming F$_{22}$ as the fundamental radial mode and F$_1$ as the first overtone \citep{2005A&A...434.1055G,1981ApJ...249..218F,2005A&A...440.1097P}. We identify F$_2$ as a non-radial mode with $l$=2. Similarly F$_3$ and F$_7$ are both found to be non-radial modes with $l$=3. Finally, F$_6$ and F$_{13}$ are also tentatively classified as non-radial modes, with possible $l$-values of 1, 2, or 3, though their exact values remain uncertain.

As shown in the top and bottom panels of Figure \ref{fig:lc}, the high pulsation amplitudes vary with time, suggesting potential modulation with the orbital period of the inner binary. Such variations, which could indicate tidally induced pulsations, have been observed in systems like HD 74423 \citep{2020NatAs...4..684H}, TIC 63328020 \citep{2021MNRAS.503..254R}, and CO Cam \citep{2020MNRAS.494.5118K}. To investigate this possibility, we analyzed the \textit{TESS} photometry to search for modulation of the pulsation signals (e.g., frequency splittings, or amplitude and phase variations phased with the orbital period). However, unlike the case of HD 74423, we did not detect any pulsation frequency splitting modulated at the orbital frequency.
\begin{figure}[ht!]
	\plotone{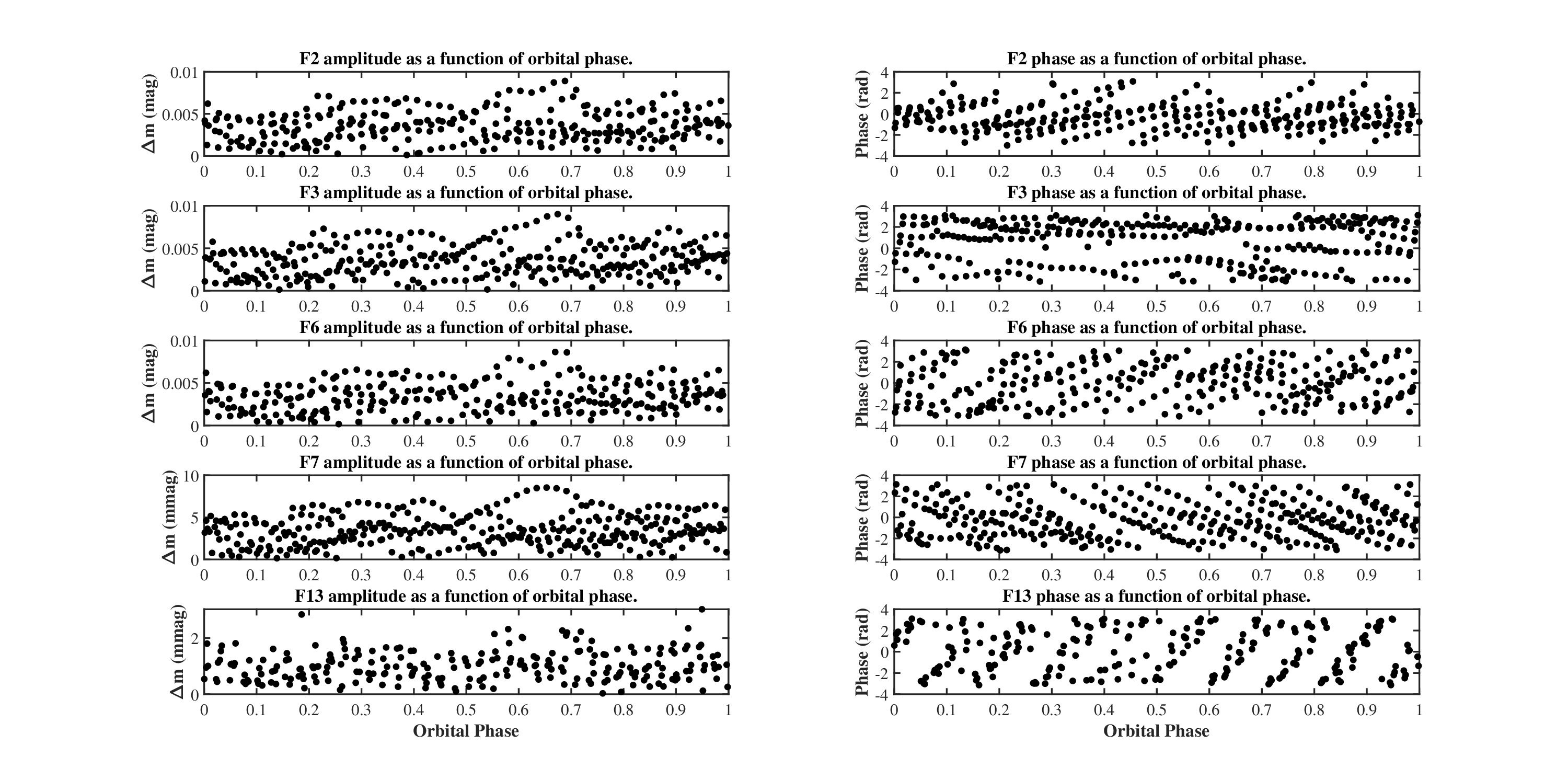}
	\caption{The five non-radial pulsation amplitudes and phases, modulated over the orbital period of the inner binary. The time zero point is set to $t_0 = \mathrm{BJD}\, 2458871.77213$, chosen such that the phases of the first two orbital sidelobes are equal.} \label{fig:Amp}
\end{figure}

Furthermore, following the method of \cite{2020NatAs...4..684H}, we performed a least-squares fit to sections of the data to examine orbital-phase-dependent variations in amplitude and phase for the non-radial modes F$_2$, F$_3$, F$_6$, F$_7$, and F$_{13}$. The data were segmented into intervals of 0.09~d (for F$_2$, F$_3$, F$_6$), 0.08~d (for F$_7$), and 0.10~d (for F$_{13}$). As shown in Figure \ref{fig:Amp}, no significant amplitude or phase modulation over the orbital cycle of the inner binary was found for these five non-radial pulsation modes. This indicates that the observed pulsations likely originate from component B rather than being tidally modulated by the inner binary, which is consistent with the conclusion of \cite{2005MNRAS.356..545F} that the pulsations arise from component B.

\section{Stellar models} \label{sec:stellar models}
\subsection{Input Physics}
We thus utilize the stellar evolution code Modules for Experiments in Stellar Astrophysics (MESA, version r22.11.1), developed by \citet{2011ApJS..192....3P,2013ApJS..208....4P,2015ApJS..220...15P,2018ApJS..234...34P}, to compute evolutionary and pulsational models for component B. Specifically, for generating stellar evolution models and calculating the adiabatic frequencies of radial and non-radial modes, we employ the \textit{pulse\_adipls} \citep{2008Ap&SS.316..113C} submodule in MESA. Our calculations adopt the 2005 update of the OPAL equation of state tables \citep{2002ApJ...576.1064R}. For opacity, we utilize the OPAL high-temperature tables from \citet{1996ApJ...464..943I} and the low-temperature tables from \citet{2005ApJ...623..585F}. We assume the initial metallicity is identical to the solar value \citep{2009ARA&A..47..481A}. Convective zones are treated using the classical mixing-length theory \citep{1958ZA.....46..108B} with a mixing-length parameter $\alpha = 1.90$ \citep{2011ApJS..192....3P}.

To model convective core overshooting, we adopt an exponentially decaying prescription for the overshooting - mixing diffusion coefficient \citep{1996A&A...313..497F, 2000A&A...360..952H}:
\begin{eqnarray}\label{equation (2)}
	\mathit{D}_{\rm ov} = \mathit{D}_0 \exp \left( \frac{-2z}{\mathit{f}_{\rm ov} \mathit{H}_{\rm p}} \right),
\end{eqnarray}
where $\mathit{D}_0$ is the diffusion coefficient at the edge of the convective core, $z$ is the distance into the radiative zone from the edge, $\mathit{f}_{\rm ov}$ is an adjustable parameter describing the efficiency of overshooting, and $\mathit{H}_{\rm p}$ is the pressure scale height. The lower limit of the diffusion coefficient is set to $\mathit{D}^{\rm limit}_{\rm ov} = 1 \times 10^{-2}$ cm$^2$ s$^{-1}$ \citep{2019ApJ...887..253C}, below which overshooting is neglected in our models. In addition, we do not include elemental diffusion, stellar rotation, and magnetic fields in the stellar structure and evolution calculations.

\subsection{Grid of Stellar Models for Component B}
Stellar evolution paths and internal structures depend critically on the initial mass $M$, initial chemical composition $(X, Y, Z)$, and the convective overshooting parameter $\mathit{f}_{\rm ov}$. Following \citet{2018MNRAS.475..981L}, we adopt the helium abundance relation $Y = 0.249 + 1.33Z$, thereby reducing the number of independent parameters to $M$, $Z$, and $\mathit{f}_{\rm ov}$. We perform a grid search over stellar masses $M$ from $2.00$ to $2.30$ M$_{\odot}$ with a step of 0.01 M$_{\odot}$ according our results of light curve modeling, and metallicities  $Z$ from $0.010$ to $0.030$ in steps of $0.001$. For the overshooting at the top of the convective core, we follow the method of \cite{2019ApJ...887..253C} to take four different cases: no overshooting ($\mathit{f}_{\rm ov} = 0$), moderate overshooting ($\mathit{f}_{\rm ov} = 0.01$), intermediate overshooting ($\mathit{f}_{\rm ov} = 0.02$), and extreme overshooting ($\mathit{f}_{\rm ov} = 0.03$). 

Each star in the grid is computed from the zero-age-main-sequence to the post-main-sequence stage ($X_c = 1\times\ 10^{-5}$). The effective temperatures of $\delta$ Scuti normally vary between 6000 K and 9800 K \citep{2010aste.book.....A}. we restrict the effective temperature of stellar models  inside this range. Luminosity and radius are constrained to $1.24 < \log (L/$L$_{\odot}) < 1.53$ and $0.39 <  \log (R/$R$_{\odot}) < 0.53 $, based on results from our light curve modeling.  As a star moves along its evolutionary track into this region, the adiabatic frequencies of the radial ($\ell$ = 0) and non-radial ($\ell$ = 1, 2, and 3) oscillations are calculated for the structure model at each stage of its evolution.

Additionally, we also include stellar rotation as a fourth adjustable parameter, varying the rotation velocity $\mathit{V}_{\rm rot}$ from 30  to 50 km s$^{-1}$ in steps of 1 km s$^{-1}$ due to the component B's project velocity is $v \sin i \geq 30$ km s$^{-1}$ \citep{2005MNRAS.356..545F}. For each pulsation mode at a given $\mathit{P}_{\rm rot}$, rotational splitting produces $2l + 1$ frequency components according to:
\begin{eqnarray}\label{equation (3)}
	\nu_{l,n,m} = \nu_{l,n} + m\delta\nu_{l,n} = \nu_{l,n} + \beta_{l,n} \frac{m}{\mathit{P}_{\rm rot}}
\end{eqnarray}
\citep{1981ApJ...244..299S,1992ApJ...394..670D,2010aste.book.....A}, where $\delta\nu_{l,n}$ is the rotational splitting frequency and $\beta_{l,n}$ determines the splitting magnitude. For uniformly rotating stars, $\beta_{l,n}$ is given by \cite{2010aste.book.....A}:
\begin{eqnarray}\label{equation (4)}
	\beta_{l,n} = \frac{\int_{0}^{R} \left( \xi^2_{r} + L^2\xi^2_{h} - 2\xi_{r}\xi_{h} - \xi^2_{h} \right) r^2 \rho  \,dr}{\int_{0}^{R} \left( \xi^2_{r} + L^2\xi^2_{h} \right) r^2 \rho \,dr},
\end{eqnarray}
where $\xi_{r}$ and $\xi_{h}$ are the radial and horizontal displacement eigenfunctions, $\rho$ is the local density, and $L^2 = l(l+1)$. According to Equation \ref{equation (3)}, each dipole mode splits into three different components, corresponding to modes with $m$ = $-$1, 0, and $+$1, respectively. Each quadrupole mode splits into five different components, corresponding to modes with $m$ = $-$2, $-$1, 0, $+$1, and $+$2, respectively.

\subsection{Fitting Results of Component B}
Based on the mode identification in section \ref{sec:fre}, we compare model frequencies with spherical degrees $l = 0$, $1$, $2$, and $3$ to the observed frequencies $F_1$, $F_2$, $F_3$, $F_6$, $F_7$, $F_{13}$, and $F_{22}$ to identify the optimal model. The goodness of fit is evaluated using:
\begin{eqnarray}\label{equation (5)}
	S^2 = \frac{1}{k} \sum_{i=1}^{k} |f_{\mathrm{model},i} - f_{\mathrm{obs},i}|^2,
\end{eqnarray}
where $f_{\mathrm{obs},i}$ is the $i$-th observed frequency, $f_{\mathrm{model},i}$ is the corresponding model frequency, and $k$ is the number of observed modes. The frequencies $F_6$ and $F_{13}$ could not be precisely identified in advance; therefore, the nearest theoretical frequencies were assigned to represent them in the model.
\begin{figure}[ht!]
	\plotone{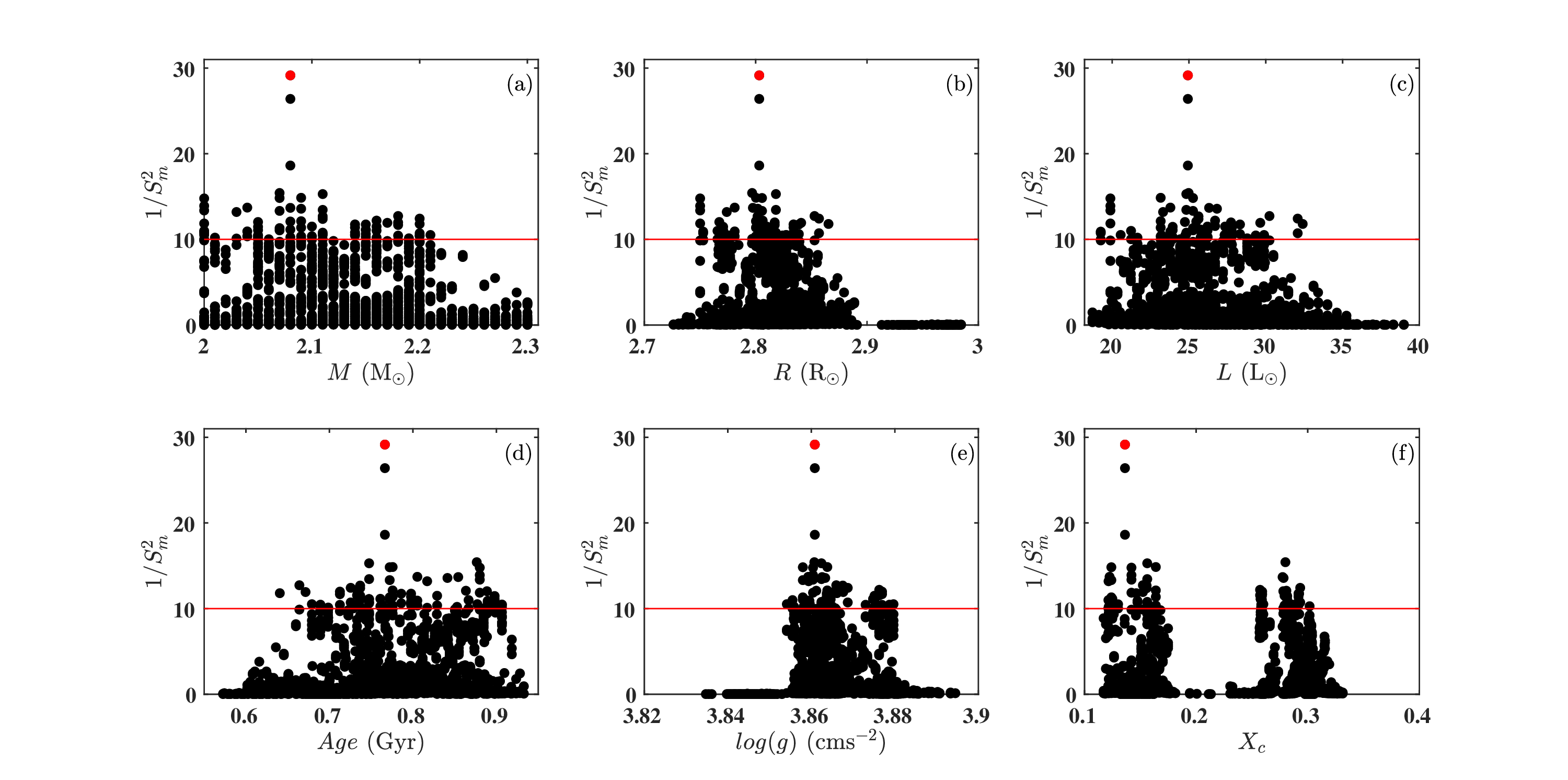}
	\caption{This visualization presents the fitting metric $S^{2}_m$ and physical parameters: stellar mass ($M$), radius ($R$), luminosity ($L$), age ($\mathrm{Age}$), gravitational acceleration ($\log g$), and central hydrogen abundance ($X_c$) with a red horizontal line marking the threshold $S^{2}_m = 0.10$, where black dots indicate the minimum $S^{2}_m$ for each model while the red dot identifies the minimum $S^{2}_m$ of the best-fitting model (Model A3).
		\label{fig:MESA1}}
\end{figure}
\begin{figure}[ht!]
	\plotone{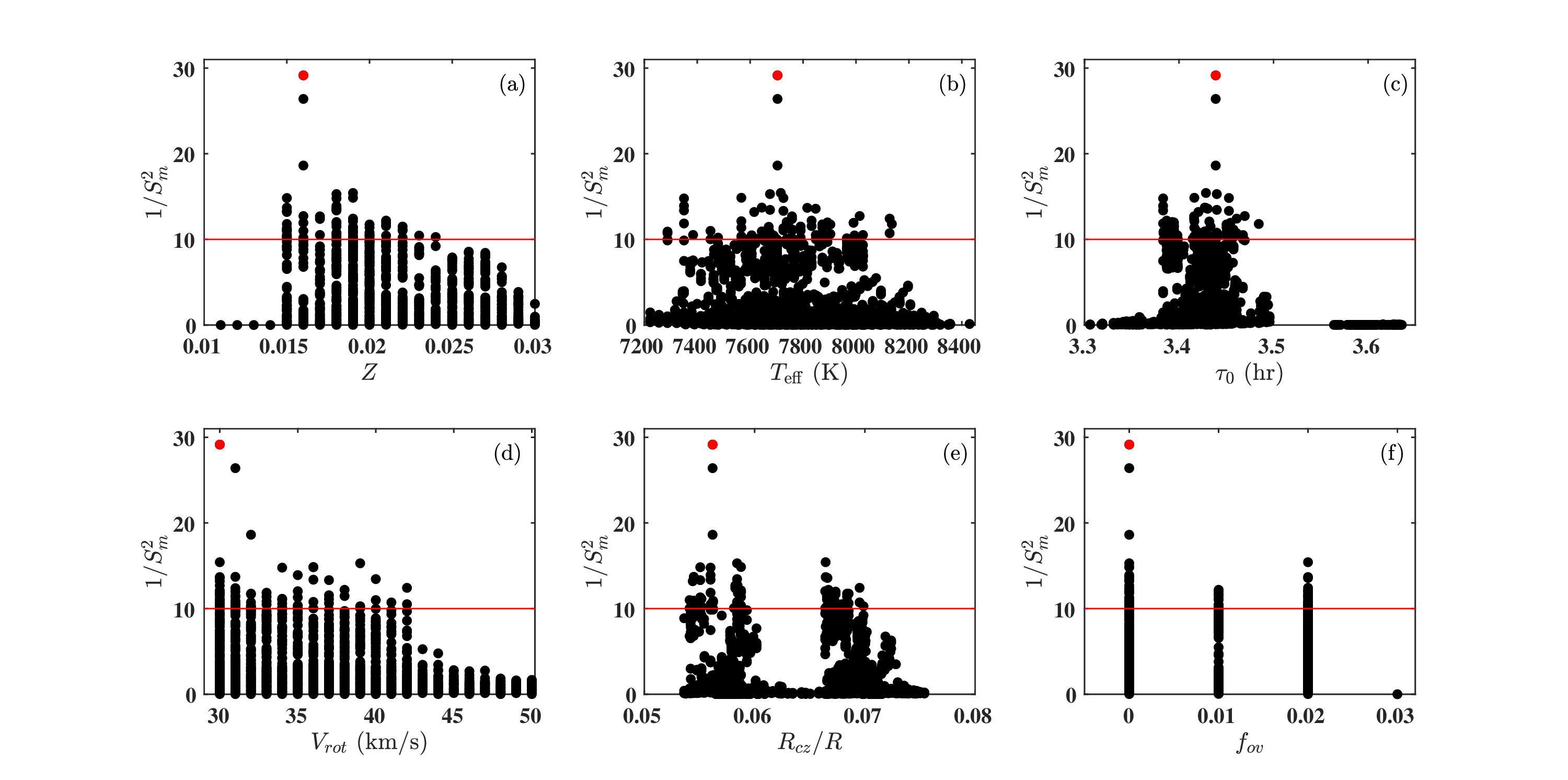}
	\caption{This visualization presents the fitting results $S^{2}_m$ alongside key physical parameters: metallicity ($Z$), effective temperature ($T_{\rm eff}$), acoustic radius ($\tau_0$), stellar rotation velocity ($V_{\rm rot}$), convective-core-to-stellar-radius ratio ($R_{\rm cz}/R$), and overshooting parameter ($f_{\rm ov}$). A red horizontal line marks the threshold $S^{2}_m = 0.10$. Black dots denote the minimum $S^{2}_m$ for each model, while a red dot identifies the minimum $S^{2}_m$ for the best-fitting model (Model A3).
		\label{fig:MESA2}}
\end{figure}

Figures \ref{fig:MESA1} and \ref{fig:MESA2} display the variation of the goodness-of-fit $S^2_{m}$ across different physical parameters.  Each dot in the figures represents one minimum value of $S_2$ along one evolutionary track, denoted as $S^2_{m}$. Horizontal lines in both figures indicate $S^2_{m} = 0.10$, corresponding to the squared value of the frequency resolution $\delta f$. Dots above this threshold correspond to 84 candidate models that are listed in Table \ref{tab:6} (see in Appendix). Model A3 achieves the global minimum $S^2_{m} = 0.0343$, establishing it as the optimal solution. This best-fitting model is highlighted by red dots in Figures \ref{fig:MESA1} and \ref{fig:MESA2}.

Figures \ref{fig:MESA1} (a)--(c) display the variation of $S^2_m$ versus stellar mass ($M$), radius ($R$), and luminosity ($L$). These parameters demonstrate excellent convergence with best-fit values $M = 2.08^{+0.13}_{-0.08}$ M$_{\odot}$, $R = 2.80^{+0.07}_{-0.05}$ R$_{\odot}$, and $L = 24.93^{+7.47}_{-5.67}$ L$_{\odot}$. Meanwhile, Figures \ref{fig:MESA1} (d)--(f) show $S^2_m$ as functions of age ($\mathrm{Age}$), gravitational acceleration ($\log g$), and central hydrogen abundance ($X_{\mathrm{c}}$). The parameters age and gravitational acceleration exhibit tight constraints: $\mathrm{Age} = 0.7664^{+0.1402}_{-0.1258}~\mathrm{Gyr}$ and $\log g = 3.8608^{+0.0189}_{-0.0067}~\mathrm{cm\,s^{-2}}$. In contrast, the central hydrogen abundance spans a broader range $X_{\mathrm{c}} = 0.1362^{+0.1655}_{-0.0148}$.

Figures \ref{fig:MESA2} (a)--(c) depict the variation of $S^2_m$ with metallicity ($Z$), effective temperature ($T_{\mathrm{eff}}$), and acoustic radius ($\tau_0$). Excellent convergence is observed for metallicity and acoustic radius ($\tau_0$) with best-fit values $Z = 0.016^{+0.008}_{-0.001}$ and $\tau_0 = 3.4388^{+0.0456}_{-0.0556}~\mathrm{hr}$, while the effective temperature shows a broader distribution $T_{\mathrm{eff}} = 7703^{+432}_{-415}~\mathrm{K}$. Meanwhile, Figures \ref{fig:MESA2} (d)--(f) present $S^2_m$ as functions of rotation velocity ($V_{\mathrm{rot}}$), convective-core-to-stellar-radius ratio ($R_{\mathrm{cz}}/R$), and overshooting parameter ($f_{\mathrm{ov}}$). All three parameters exhibit good convergence: $V_{\mathrm{rot}} = 30^{+12}_{-0}~\mathrm{km\,s^{-1}}$, $R_{\mathrm{cz}}/R = 0.0562^{+0.0137}_{-0.0021}$, and $f_{\mathrm{ov}} = 0.00-0.02$.

Table~\ref{tab:4} presents a comparison between the observed frequencies and those of the best-fitting model (Model A3). Based on this comparison, $F_{22}$ and $F_1$ are identified as the fundamental radial mode and the first radial overtone, respectively. $F_2$ is classified as a quadrupole mode, while $F_3$, $F_6$, and $F_7$ are recognized as octupole modes. Although $F_{13}$ does not match any f-mode predicted by the model, it coincides with a g-mode of degree $l=3$. We therefore identify $F_{13}$ as a low-order g-mode with $l=3$. The asteroseismically derived parameters of component~B, which are in good agreement with the light curve modeling results, are listed in Column~2 of Table~\ref{tab:5}. These non-radial modes provide crucial constraints on the internal structure of component~B in DG~Leo.
\begin{deluxetable*}{lcccc}  
	\tabletypesize{\scriptsize}  
	\tablecaption{Comparisons between model frequencies of best-fitting model (Model A3) and observations.\label{tab:4}}  
	\tablehead{  
		\colhead{ID} & \colhead{$F_{\rm obs}$ ($\mu$Hz)} & \colhead{$F_{\rm mod}$ ($\mu$Hz)} & \colhead{$(\ell,\,n_p,\,n_g,\,m)$} & \colhead{$|F_{\rm obs} - F_{\rm mod}|$ ($\mu$Hz)}  
	}  
	\startdata  
	$F_{22}$  & 109.4199 & 109.3071 & $(0,\,0,\,0,\,0)$   & 0.1128 \\  
	$F_1$  & 140.1847 & 139.7799 & $(0,\,1,\,0,\,0)$  & 0.4048 \\  
	$F_2$  & 131.3715 & 131.2754 & $(2,\,1,\,-3,\,1)$   & 0.0961\\  
	$F_7$  & 147.5550 & 147.4214 & $(3,\,1,\,-3,\,0)$  & 0.1336 \\  
	$F_6$& 125.3907 & 125.5038 & $(3,\,1,\,-5,\,2)$   & 0.1131 \\
	$F_3$& 138.8162 & 138.6726 & $(3,\,1,\,-4,\,-1)$   & 0.1436 \\
	$F_{13}$& 106.2529 & 106.3081 & $(3,\,0,\,-6,\,-2)$   & 0.0481 \\  
	\tableline  
	\enddata  
	\tablecomments{$F_{\rm obs}$ is the observed frequency. $F_{\rm mod}$ is the model frequency. ($\ell$, $n_p$, $n_g$, $m$) are the spherical harmonic degree, the radial orders in
			the p-mode propagation zone, the radial orders in the g-mode propagation zone, and the azimuthal number of themodel frequency.}
\end{deluxetable*}
\begin{deluxetable*}{lcc}  % 列格式：参数列左对齐(l) + 两模型列居中(cc)
	\tabletypesize{\scriptsize}  % 小字号适配页面
	\tablecaption{ Fundamental parameters of the component B of DG Leo.\label{tab:5}}
	\tablehead{  % 表头三列：参数、单星模型、吸积模型
		\colhead{Parameters} & \colhead{Single-star Models}
	}
	\startdata  % 数据区开始
	$Z$ & $0.015$--$0.024$ ($0.016^{+0.008}_{-0.001}$)  \\
	$M$ (M$_\odot$) & $2.00$--$2.21$ ($2.08^{+0.13}_{-0.08}$) \\
	$V_{\rm rot}$ (kms$^{-1}$) & $30$--$42$ ($30^{+12}_{-0}$)\\
	$T_{\rm eff}$ (K) & $7287$--$8135$ ($7703^{+432}_{-415}$)\\
	$\log(g)$ (cms$^{-2}$) & $3.8541$--$3.8797$ ($3.8608^{+0.0189}_{-0.0067}$) \\
	$R$ (R$_\odot$) & $2.75$--$2.87$ ($2.80^{+0.07}_{-0.05}$) \\
	$L$ (L$_\odot$) & $19.26$--$32.40$ ($24.93^{+7.47}_{-5.67}$) \\
	$\tau_0$ (hr) & $3.3832$--$3.4844$ ($3.4388^{+0.0456}_{-0.0556}$) \\
	$X_c$ & $0.1214$--$0.3017$ ($0.0.1362^{+0.1655}_{-0.0148}$)\\
	$R_{cz}/R$ & $0.0541$--$0.0699$ ($0.0562^{+0.0137}_{-0.0021}$)\\
	$Age$ (Gyr) & $0.6406$--$0.9066$ ($0.7664^{+0.1402}_{-0.1258}$) \\
	\tableline  % deluxetable专用横线（替代\hline）
	\enddata  % 数据区结束
	
	\tablecomments{$V_{\rm rot}$ denotes the rotation velocity. $\tau_0$ is the acoustic radius. $X_c$ is the central hydrogen abundance.}
	
\end{deluxetable*}

\section{Discussion and conclusions}\label{sec:dc}
The hierarchical and spectroscopic triple system DG~Leo, consisting of an inner binary and a distant $\delta$ Scuti star, exhibits both ellipsoidal variations and high-amplitude pulsations in its \textit{TESS} light curve. Through modelling of these ellipsoidal variations, our analysis constrains the inner binary's orbital inclination to $i = 66.87 \pm 0.60^\circ$, which is consistent with prior rough estimates \citep{1977PASP...89..658B,1991PASP..103..628R,2005MNRAS.356..545F}. The other orbital parameters are also determined, too. Subsequent in combination our light curve modeling results with spectroscopic results \citep{2004RMxAC..21...73L,2005MNRAS.356..545F} yielded dynamically derived stellar parameters: Component Aa has mass $M_1 = 2.258 \pm 0.033$ M$_{\odot}$, radius $R_1 = 3.353 \pm 0.015$ R$_{\odot}$, and luminosity $L_1 = 25.99 \pm 1.44$ L$_{\odot}$; Component Ab shows $M_2 = 2.258 \pm 0.033$ M$_{\odot}$, $R_2 = 3.263 \pm 0.025$ R$_{\odot}$, $L_2 = 24.48 \pm 1.67$ L$_{\odot}$; while component B exhibits $M_B = 2.39 \pm 0.40$ M$_{\odot}$, $R_B = 2.95 \pm 0.50$ R$_{\odot}$, $L_B = 26.05 \pm 8.42$ L$_{\odot}$. When compared with the results obtained using the evolutionary method described in \cite{2005MNRAS.356..545F}, the radii ( $R_1 = 2.96 \pm 0.2$ R$_{\odot}$, $R_2 = 2.94 \pm 0.2$ R$_{\odot}$) of the two components of the inner binary are similar to our results, but the masses ($M_1 = 2.0 \pm 0.2$ M$_{\odot}$, $M_2 = 2.0 \pm 0.2$ M$_{\odot}$) are somewhat smaller.  The combination of light curve modeling and spectroscopic results reveals further: i) the inner binary is a detached system comprising two late A-type components that have low Roche lobe filling factors and similar physical parameters; ii) the parameters of component B are slightly larger than those of the inner binary system, and are consistent with them within uncertainties. Therefore, it may be concluded that the three stars in DG Leo have similar physical parameters, indicating that they are at a similar evolutionary stage and have undergone coeval evolution.
\begin{figure}[ht!]
	\centering
	\includegraphics[width=0.70\columnwidth]{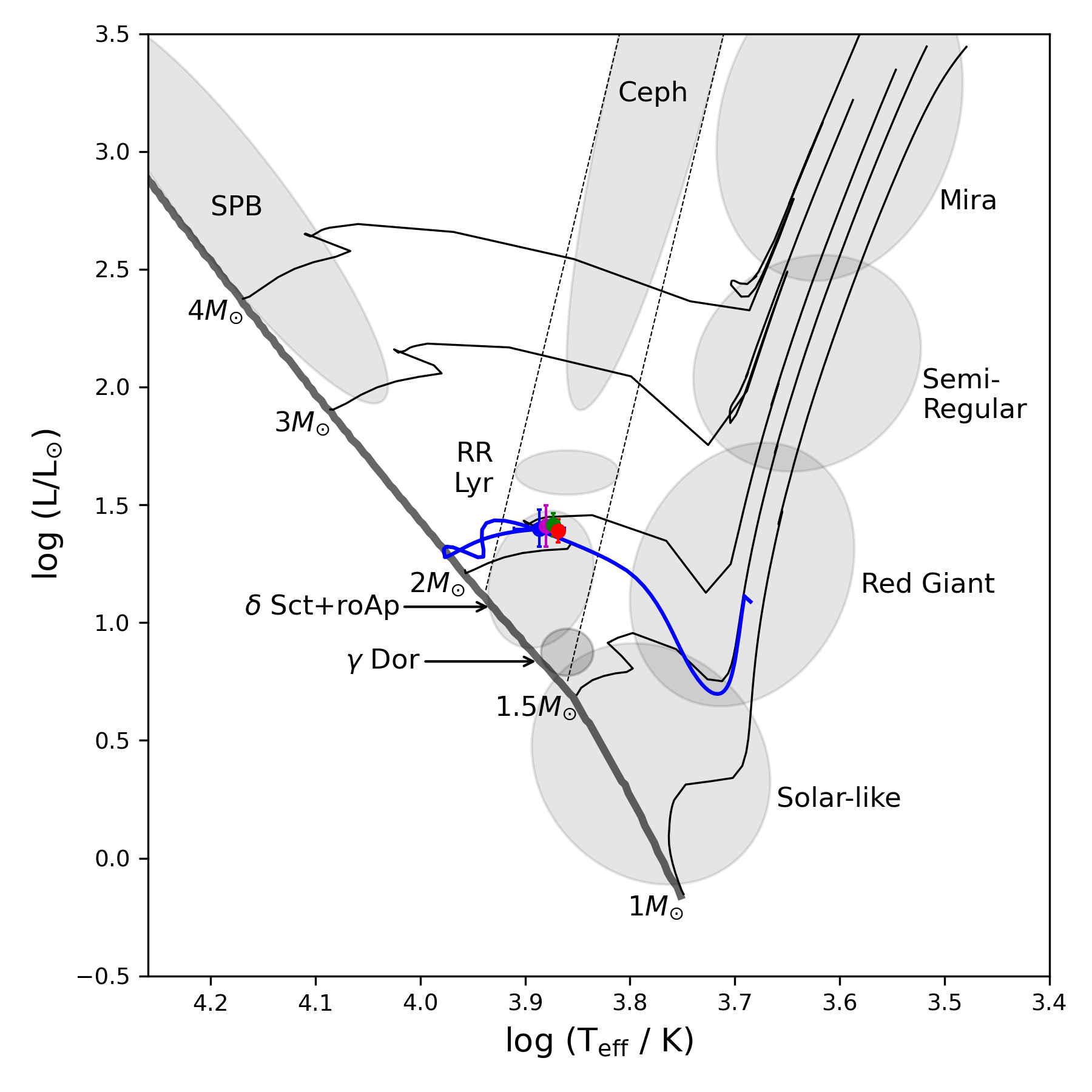} % 缩小到原宽的77%
	\caption{The H-R diagram for DG~Leo displays the positions of all three stellar components. The inner binary components are represented by a green dot (Aa) and a red dot (Ab), respectively. Component B's location is indicated by a pink dot, with all positions derived from the light curve modeling. Additionally, a blue dot denotes the current evolutionary state of component B according to the best-fitting stellar model (Model A3), while the accompanying curve traces its predicted evolutionary track.}
	\label{fig:DGHR}
\end{figure}

Moreover, no shallow eclipses of the inner binary system were detected in earlier studies \citep{2004RMxAC..21...73L,2005MNRAS.356..545F}. We also carefully examined the \textit{TESS} light curves from all four sectors and confirm the absence of any eclipses. This non-detection is well explained by the orbital geometry. The mean orbital inclination obtained from our light curve modeling (see Table~\ref{tab:solutions}) is just below the critical inclination of $68.2^\circ$ required for eclipses to occur, which is estimated from the stellar radii and orbital separation via $\cos i = \frac{R_1 + R_2}{a}$. Given that our measured inclination lies slightly below this theoretical threshold, any eclipses would be extremely shallow and undetectable amid the dominant pulsation signal.

After isolating the intrinsic pulsation signal by subtracting the binary model from the original \textit{TESS} light curve, we performed a multiple frequency analysis on the residuals. This revealed seven significant independent frequencies: $F_1$, $F_2$, $F_3$, $F_6$, $F_7$, $F_{13}$, and $F_{22}$. To identify their pulsation modes, we computed the pulsation constant $Q$ for each frequency using the standard relation. Comparison with theoretical models \citep{1981ApJ...249..218F} indicates that $F_1$ and $F_{22}$ are the first-overtone and fundamental radial modes, respectively — a classification further supported by their period ratio of 0.780. Frequencies $F_2$, $F_3$, and $F_7$ are identified as non-radial modes with $l = 2$, $3$, and $3$, respectively, while $F_6$ and $F_{13}$ are also likely non-radial. Although amplitude variations are present, a search for tidally induced modulations (e.g., frequency splitting or orbital-phase-dependent variations in amplitude or phase) yielded no significant detections. This confirms that the observed pulsations are intrinsic to component B, consistent with \citet{2005MNRAS.356..545F}.

To further verify these mode identifications, trace the evolutionary history of the $\delta$ Scuti star, and constrain the evolutionary stage of the triple system, we constructed a grid of theoretical models incorporating the dynamically measured parameters. Our asteroseismic modeling confirms $F_{22}$ and $F_1$ as the fundamental and first-overtone radial modes, respectively; $F_2$ as a quadrupole ($\ell = 2$) mode; and $F_3$, $F_6$, $F_7$, and $F_{13}$ as octupole ($\ell = 3$) modes—consistent with the earlier identification. The best-fitting model yields the following fundamental parameters for component B: $M = 2.08^{+0.13}_{-0.08}~M_{\odot}$, $R = 2.80 ^{+0.06}_{-0.05}~R_{\odot}$, $L = 24.92 ^{+7.47}_{-5.67}~L_{\odot}$, $T_{\rm eff} = 7703^{+432}_{-415}~\mathrm{K}$, and $v \sin i = 30^{+12}_{-0}~\mathrm{km~s^{-1}}$. These asteroseismic results are consistent with both our dynamical solutions and previously reported values for the effective temperature and projected rotation velocity of component B \citep{2005MNRAS.356..545F,2004ASPC..318..342F}.
\begin{figure}[ht!]
	\plotone{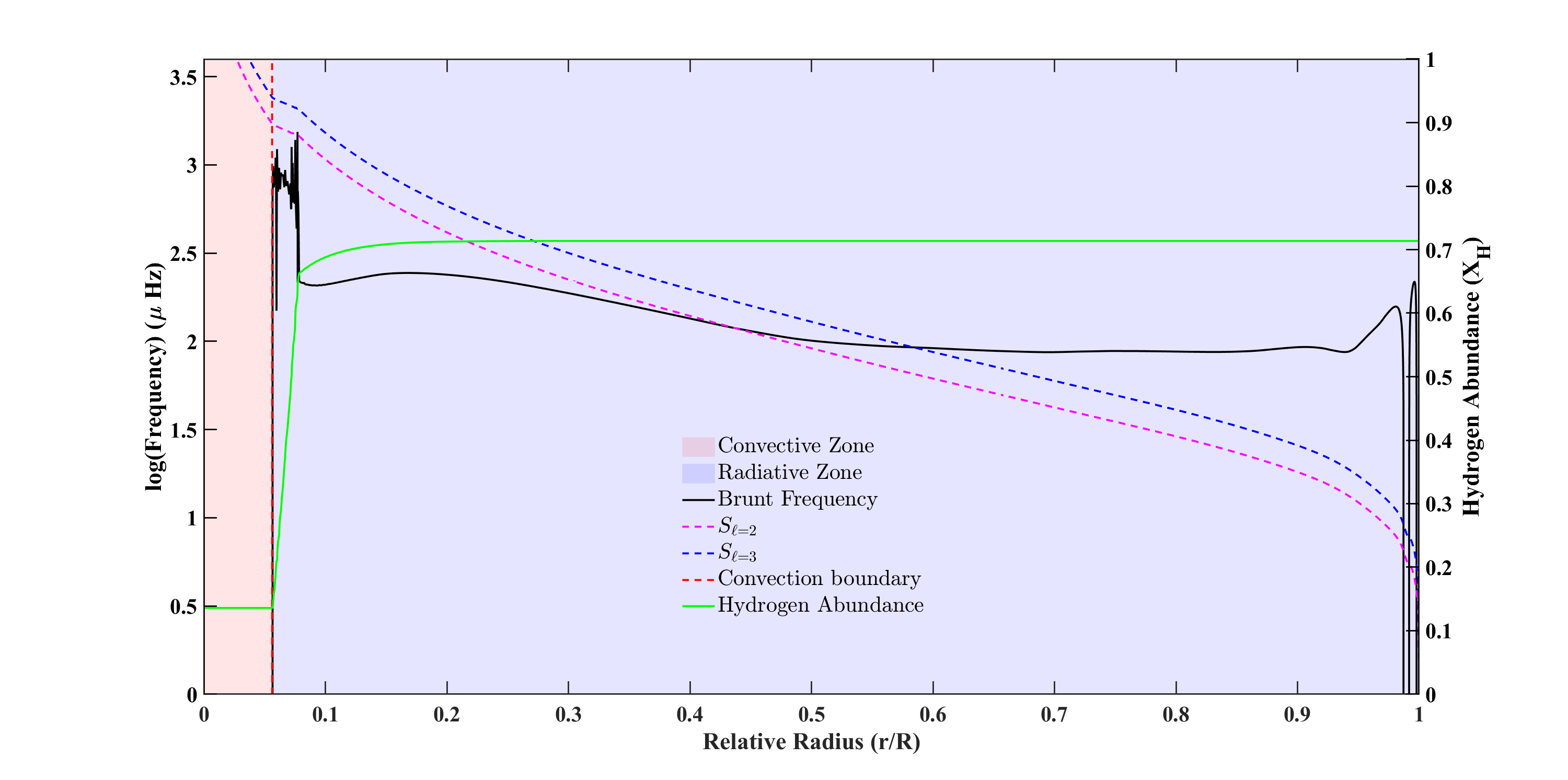}
	\caption{Visualization of the Brunt–V\"{a}is\"{a}l\"{a} frequency ($N$), characteristic acoustic oscillation frequencies ($S_\ell$ for $\ell = 2,3$), and hydrogen abundance ($X_{\mathrm{H}}$) inside the component B. The vertical dashed line indicates the boundary of convective zone  ($\nabla_r=\nabla_{ad}$).\label{fig:Brunt}}
\end{figure}

Figure~\ref{fig:DGHR} presents the H-R diagram for DG~Leo, showing positions of all three stellar components. The components of the inner binary, Aa and Ab, are marked by green and red dots, respectively, while component B is indicated by a pink dot. All positions are derived dynamically based on light curve modeling. The blue dot indicates the current evolutionary state of component B derived from theoretical models, consistent with the pink dot within uncertainties. The accompanying blue curve shows its evolutionary track from the pre-main-sequence to the post-main-sequence phase. Component B is thus confirmed to be a post-main-sequence star with an age of $0.7664^{+0.1402}_{-0.1258}$~Gyr. Moreover, we found that components Aa and Ab have evolved slightly more than component B, but they occupy similar positions to each other in the H-R diagram within uncertainties. The near-identical positions of all three components in the diagram, closely coupled with their similar physical properties (e.g., mass and radius), support their simultaneous formation from the same protostellar environment. Therefore, we conclude that all three components are post-main-sequence stars that share the same age and evolutionary stage. Component B exhibits several characteristics of HADS variables: it is located near the HADS region in the H--R diagram, pulsates predominantly in the first-overtone radial mode, and has a spectral type consistent with late-A type stars \citep{2005MNRAS.356..545F}. However, its observed amplitude is only 0.01 mag (Figure~\ref{fig:lc})—approximately one-tenth of the typical amplitude for HADS stars—likely due to photometric dilution from the brighter inner binary. We suggest that the component B may be an HADS-like star. Furthermore, the reason for the low amplitude observed in the fundamental mode ($F_{22}$) remains unclear.

\begin{figure}[ht!]
	\plotone{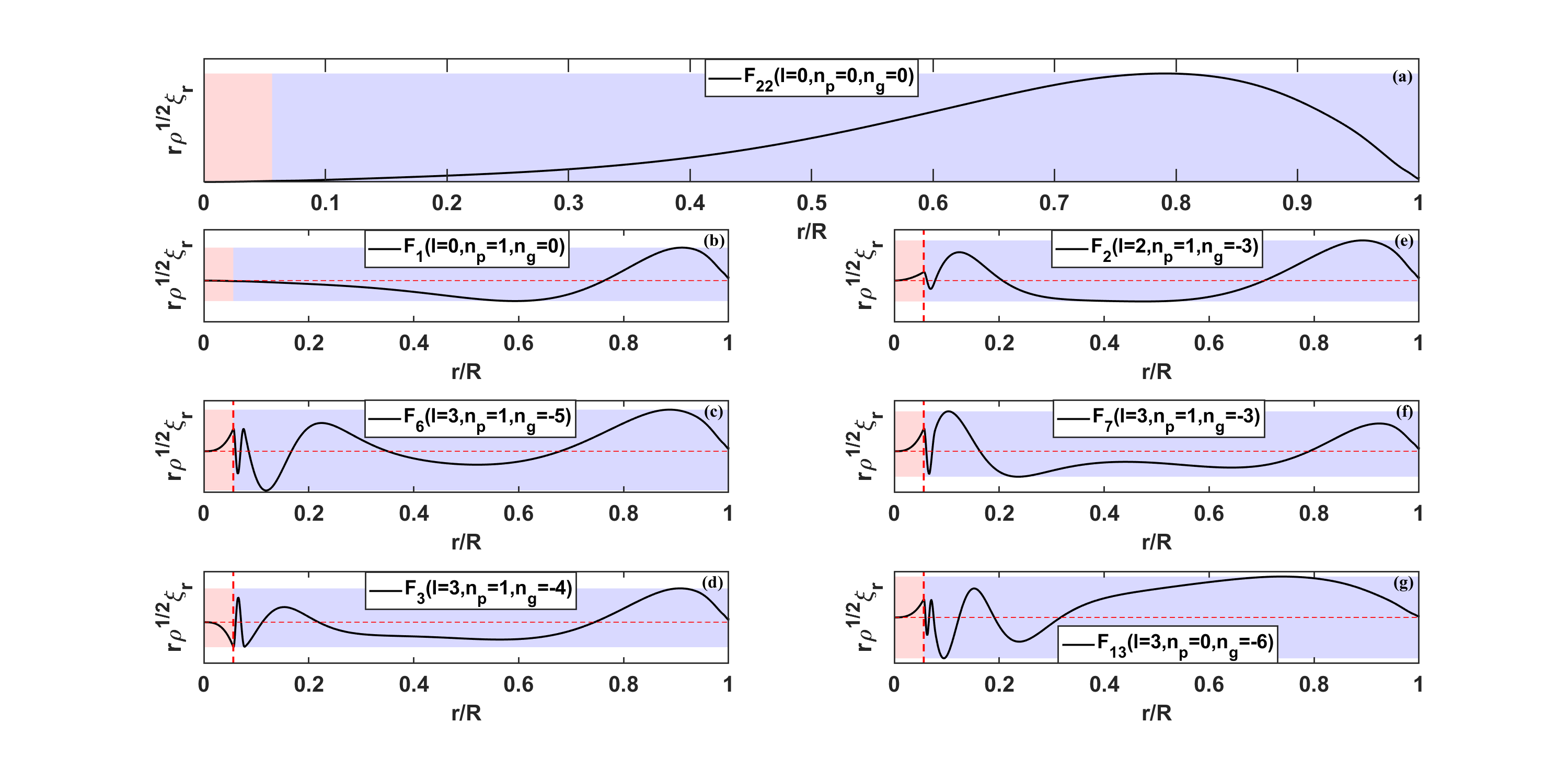}
	\caption{ Scaled radial displacement eigenfunctions corresponding to the observed frequencies in the best-fitting model. Panel (a) corresponds to the fundamental radial mode $F_{22}$; panel (b) to the first radial overtone mode $F_1$; panels (c), (d), (e), (f), and (g) correspond to the observed frequencies $F_6$, $F_3$, $F_2$, $F_7$, and $F_{13}$, respectively. The vertical red dashed lines mark the boundary of the convective core ($\nabla_r=\nabla_{ad}$).  \label{fig:Eigen}}
\end{figure}

Figure~\ref{fig:Brunt} displays radial profiles of the Brunt–V\"{a}is\"{a}l\"{a} frequency ($N$), characteristic acoustic frequencies ($S_\ell$ for $\ell = 2,3$), and hydrogen abundance ($X_{\mathrm{H}}$) in the best-fitting stellar model. Figure~\ref{fig:Eigen} reveals that the fundamental radial mode $F_{22}$ and radial first overtone $F_1$ propagate primarily in the stellar envelope, thereby probing envelope structure. It is also evident that substantial amplitudes of non-radial oscillation modes are present near both the convective core boundary and the envelope. This indicates that these modes exhibit a mixed character. Consequently, non-radial modes provide strong constraints when determining the convective core boundary.

The acoustic radius $\tau_0$ (sound travel time from center to surface) is defined as \citep{2010aste.book.....A}:
\begin{equation}\label{eq:acoustic_radius}
	\tau_0 = \int_{0}^{R} \frac{dr}{c_s}
\end{equation}
where $\mathrm{c}_s$ denotes the adiabatic sound speed. Since $c_s$ in the stellar envelope is significantly lower than in the convective core, $\tau_0$ predominantly characterizes envelope properties. Given that the non-radial modes have substantial amplitudes near the edge of the convective core, we utilize the relative core radius $R_{\mathrm{cz}}/R$ to describe the features of the deep interior structure of component B. Consequently, fitting the seven $\delta$ Scuti frequencies requires matching both the convective core and stellar envelope in theoretical models to the actual structure of the component B. Figures \ref{fig:MESA2} shows that $\tau_0$ of the candidate models converges well to 3.4388$^{+0.0456}_{-0.0556}$ hr and $R_{\mathrm{cz}}/R$ of the candidate models converges well to 0.0562$^{+0.0137}_{-0.0021}$. This suggests that they are nearly alike in structure.

In summary, DG Leo establishes a hierarchical triple system comprising an inner spectroscopic binary (components Aa and Ab) and a distant component B. All three stars have reached the post-main-sequence stage. This unique configuration provides a natural laboratory for probing coeval stellar evolution, asteroseismology, and dynamical interactions in triple systems. Further investigation of the orbital motion of component B necessitates more precise astrometry from Gaia \citep{2016A&A...595A...1G,2018A&A...616A...1G} to constrain its long-period orbit. Meanwhile, additional medium- or high-resolution spectral data are essential for resolving non-radial pulsation modes and elemental abundances, as well as for determining more precise physical parameters for component B in the future. This would advance stellar structure and evolution models for DG Leo.

%% Also note that the akcnowlodgment environment does not support long amounts of text. If you have a lot of people and institutions to acknowledge, do not use this command. Instead, create a new \section{Acknowledgments}.
\begin{acknowledgments}
	This work is supported by the International Cooperation Projects of the National Key R\&D  Program (No. 2022YFE0127300), the 2022 CAS “Light of West China” Program, the National Natural Science Foundation of China (No. 12573038), the Postdoctoral Fellowship Program of CPSF under Grant Number GZC20252095, the China Postdoctoral Science Foundation under Grant Number 2025M773194, and the Caiyun Postdoctoral Program in Yunnan Province of China (grant No. C615300504124). The \textit{TESS} data presented in this paper were obtained from the Mikulski Archive for Space Telescopes (MAST) at the Space Telescope Science Institute (STScI). STScI is operated by the Association of Universities for Research in Astronomy, Inc. Support to MAST for these data is provided by the NASA Office of Space Science. Funding for the \textit{TESS} mission is provided by the NASA Explorer Program. The authors are sincerely grateful to an
	anonymous referee for instructive advice and productive suggestions.
\end{acknowledgments}


%%\vspace{5mm}

%%\appendix

\section{Appendix information}
In this Appendix, we present the 84 candidate models generated using the MESA code (see Section \ref{sec:stellar models} for details). The physical parameters of these models—including stellar age, effective temperature, mass, central hydrogen abundance, gravitational acceleration, stellar luminosity, stellar radius, metallicity, stellar rotation velocity, acoustic radius, and convective-core-to-stellar-radius ratio—are listed in Table \ref{tab:6}. Among these candidates, Model A3 achieves the global minimum of $S^2_{m} = 0.0343$, establishing it as the best-fitting model, which is highlighted by the red dots in Figures \ref{fig:MESA1} and \ref{fig:MESA2}.

\setlength{\tabcolsep}{3pt}
\renewcommand{\arraystretch}{0.9}
\footnotesize 
\begin{longtable}{lccccccccccccc}
	\caption{Candidate models with $S^2_m < 0.10$.\label{tab:6}} \\
	\hline
	\multicolumn{1}{c}{Mode ID} & \multicolumn{1}{c}{Age (Gyr)} & \multicolumn{1}{c}{$T_{\rm eff}$ (K)} & \multicolumn{1}{c}{$M$ (M$_\odot$)} & \multicolumn{1}{c}{$X_c$} & \multicolumn{1}{c}{log $g$ (cms$^{-2}$)} & \multicolumn{1}{c}{$L$ (L$_\odot$)} & \multicolumn{1}{c}{$R$ (R$_\odot$)} & \multicolumn{1}{c}{$Z$} & \multicolumn{1}{c}{$V_{\rm rot}$ (km/s)} & \multicolumn{1}{c}{$\tau_0$ (hr)} &\multicolumn{1}{c}{$R_{cz}/R$} & \multicolumn{1}{c}{$S^2_m$} \\
	\hline
	\endfirsthead
	
	\multicolumn{13}{c}%
	{{\bfseries \tablename\ \thetable{} -- continued}} \\
	\hline
	\multicolumn{1}{c}{Mode ID} & \multicolumn{1}{c}{Age (Gyr)} & \multicolumn{1}{c}{$T_{\rm eff}$ (K)} & \multicolumn{1}{c}{$M$ (M$_\odot$)} & \multicolumn{1}{c}{$X_c$} & \multicolumn{1}{c}{log $g$ (cms$^{-2}$)} & \multicolumn{1}{c}{$L$ (L$_\odot$)} & \multicolumn{1}{c}{$R$ (R$_\odot$)} & \multicolumn{1}{c}{$Z$} & \multicolumn{1}{c}{$V_{\rm rot}$ (km/s)} & \multicolumn{1}{c}{$\tau_0$ (hr)} &\multicolumn{1}{c}{$R_{cz}/R$} & \multicolumn{1}{c}{$S^2_m$} \\
	\hline
	\endhead
	
	\hline \multicolumn{13}{r}{{Continued}} \\
	\endfoot
	
	\hline
	\multicolumn{13}{l}{{\small $V_{\rm rot}$ is the rotation velocity}. $\tau_0$ is the acoustic radius. $X_c$ is the central hydrogen abundance.}\\
	\endlastfoot
	
	A1   & 0.7664 & 7703 & 2.08 & 0.1362 & 3.8608 & 24.93 & 2.80 & 0.016 & 31.00 & 3.4388 & 0.0562 & 0.0379 \\
	A2   & 0.7664 & 7703 & 2.08 & 0.1362 & 3.8608 & 24.93 & 2.80 & 0.016 & 32.00 & 3.4388 & 0.0562 & 0.0537 \\
	\textbf{A3}   & \textbf{0.7664} & \textbf{7703} & \textbf{2.08} & \textbf{0.1362} & \textbf{3.8608} & \textbf{24.93} & \textbf{2.80} & \textbf{0.016} & \textbf{30.00} & \textbf{3.4388} & \textbf{0.0562} & \textbf{0.0343} \\
	A4   & 0.7285 & 7848 & 2.11 & 0.1301 & 3.8596 & 27.32 & 2.83 & 0.015 & 31.00 & 3.4677 & 0.0560 & 0.0952 \\
	A5   & 0.7285 & 7848 & 2.11 & 0.1301 & 3.8596 & 27.32 & 2.83 & 0.015 & 30.00 & 3.4677 & 0.0560 & 0.0998 \\
	A6   & 0.6406 & 8135 & 2.20 & 0.1463 & 3.8661 & 32.40 & 2.87 & 0.015 & 30.00 & 3.4844 & 0.0585 & 0.0847 \\
	A7   & 0.7368 & 7758 & 2.11 & 0.1556 & 3.8675 & 25.62 & 2.80 & 0.017 & 30.00 & 3.4159 & 0.0583 & 0.0788 \\
	A8   & 0.7368 & 7758 & 2.11 & 0.1556 & 3.8675 & 25.62 & 2.80 & 0.017 & 31.00 & 3.4159 & 0.0583 & 0.0807 \\
	A9   & 0.7610 & 7755 & 2.08 & 0.1249 & 3.8582 & 25.76 & 2.81 & 0.015 & 35.00 & 3.4577 & 0.0552 & 0.0975 \\
	A10  & 0.7610 & 7755 & 2.08 & 0.1249 & 3.8582 & 25.76 & 2.81 & 0.015 & 34.00 & 3.4577 & 0.0552 & 0.0897 \\
	A11  & 0.6640 & 8014 & 2.18 & 0.1523 & 3.8660 & 30.24 & 2.85 & 0.016 & 30.00 & 3.4696 & 0.0584 & 0.0786 \\
	A12  & 0.6714 & 7991 & 2.17 & 0.1523 & 3.8676 & 29.64 & 2.84 & 0.016 & 30.00 & 3.4543 & 0.0588 & 0.0838 \\
	A13  & 0.7722 & 7726 & 2.07 & 0.1240 & 3.8580 & 25.26 & 2.81 & 0.015 & 36.00 & 3.4528 & 0.0551 & 0.0674 \\
	A14  & 0.7722 & 7726 & 2.07 & 0.1240 & 3.8580 & 25.26 & 2.81 & 0.015 & 34.00 & 3.4528 & 0.0551 & 0.0905 \\
	A15  & 0.7722 & 7726 & 2.07 & 0.1240 & 3.8580 & 25.26 & 2.81 & 0.015 & 37.00 & 3.4528 & 0.0551 & 0.0750 \\
	A16  & 0.7722 & 7726 & 2.07 & 0.1240 & 3.8580 & 25.26 & 2.81 & 0.015 & 35.00 & 3.4528 & 0.0551 & 0.0884 \\
	A17  & 0.8771 & 7384 & 2.00 & 0.1275 & 3.8541 & 20.55 & 2.77 & 0.017 & 30.00 & 3.4215 & 0.0545 & 0.0951 \\
	A18  & 0.8529 & 7479 & 2.01 & 0.1217 & 3.8559 & 21.65 & 2.77 & 0.016 & 30.00 & 3.4225 & 0.0542 & 0.0980 \\
	A19  & 0.8053 & 7644 & 2.04 & 0.1224 & 3.8593 & 23.79 & 2.78 & 0.015 & 32.00 & 3.4272 & 0.0545 & 0.0966 \\
	A20  & 0.8053 & 7644 & 2.04 & 0.1224 & 3.8593 & 23.79 & 2.78 & 0.015 & 30.00 & 3.4272 & 0.0545 & 0.0915 \\
	A21  & 0.8053 & 7644 & 2.04 & 0.1224 & 3.8593 & 23.79 & 2.78 & 0.015 & 31.00 & 3.4272 & 0.0545 & 0.0729 \\
	A22  & 0.8170 & 7615 & 2.03 & 0.1215 & 3.8594 & 23.31 & 2.77 & 0.015 & 31.00 & 3.4205 & 0.0544 & 0.0990 \\
	A23  & 0.8170 & 7615 & 2.03 & 0.1215 & 3.8594 & 23.31 & 2.77 & 0.015 & 30.00 & 3.4205 & 0.0544 & 0.0758 \\
	A24  & 0.8667 & 7450 & 2.00 & 0.1214 & 3.8555 & 21.23 & 2.77 & 0.016 & 40.00 & 3.4184 & 0.0541 & 0.0910 \\
	A25  & 0.8667 & 7450 & 2.00 & 0.1214 & 3.8555 & 21.23 & 2.77 & 0.016 & 39.00 & 3.4184 & 0.0541 & 0.0998 \\
	A26  & 0.8896 & 7287 & 2.00 & 0.1474 & 3.8595 & 19.26 & 2.75 & 0.019 & 32.00 & 3.3836 & 0.0562 & 0.0932 \\
	A27  & 0.8896 & 7287 & 2.00 & 0.1474 & 3.8595 & 19.26 & 2.75 & 0.019 & 31.00 & 3.3836 & 0.0562 & 0.0917 \\
	A28  & 0.8896 & 7287 & 2.00 & 0.1474 & 3.8595 & 19.26 & 2.75 & 0.019 & 30.00 & 3.3836 & 0.0562 & 0.0934 \\
	A29  & 0.8896 & 7287 & 2.00 & 0.1474 & 3.8595 & 19.26 & 2.75 & 0.019 & 33.00 & 3.3836 & 0.0562 & 0.0921 \\
	A30  & 0.7670 & 7628 & 2.09 & 0.1571 & 3.8649 & 23.86 & 2.80 & 0.018 & 30.00 & 3.4144 & 0.0581 & 0.0994 \\
	A31  & 0.8801 & 7350 & 2.00 & 0.1420 & 3.8604 & 19.88 & 2.75 & 0.018 & 33.00 & 3.3832 & 0.0560 & 0.0843 \\
	A32  & 0.8801 & 7350 & 2.00 & 0.1420 & 3.8604 & 19.88 & 2.75 & 0.018 & 36.00 & 3.3832 & 0.0560 & 0.0748 \\
	A33  & 0.8801 & 7350 & 2.00 & 0.1420 & 3.8604 & 19.88 & 2.75 & 0.018 & 35.00 & 3.3832 & 0.0560 & 0.0719 \\
	A34  & 0.8801 & 7350 & 2.00 & 0.1420 & 3.8604 & 19.88 & 2.75 & 0.018 & 34.00 & 3.3832 & 0.0560 & 0.0677 \\
	A35  & 0.7757 & 7567 & 2.09 & 0.1639 & 3.8638 & 23.16 & 2.80 & 0.019 & 31.00 & 3.4162 & 0.0588 & 0.0879 \\
	A36  & 0.7757 & 7567 & 2.09 & 0.1639 & 3.8638 & 23.16 & 2.80 & 0.019 & 35.00 & 3.4162 & 0.0588 & 0.0965 \\
	A37  & 0.7757 & 7567 & 2.09 & 0.1639 & 3.8638 & 23.16 & 2.80 & 0.019 & 30.00 & 3.4162 & 0.0588 & 0.0825 \\
	A38  & 0.7757 & 7567 & 2.09 & 0.1639 & 3.8638 & 23.16 & 2.80 & 0.019 & 36.00 & 3.4162 & 0.0588 & 0.0673 \\
	A39  & 0.7475 & 7675 & 2.11 & 0.1562 & 3.8624 & 24.83 & 2.82 & 0.018 & 39.00 & 3.4400 & 0.0584 & 0.0654 \\
	A40  & 0.7475 & 7675 & 2.11 & 0.1562 & 3.8624 & 24.83 & 2.82 & 0.018 & 36.00 & 3.4400 & 0.0584 & 0.0927 \\
	A41  & 0.7475 & 7675 & 2.11 & 0.1562 & 3.8624 & 24.83 & 2.82 & 0.018 & 40.00 & 3.4400 & 0.0584 & 0.0744 \\
	A42  & 0.7475 & 7675 & 2.11 & 0.1562 & 3.8624 & 24.83 & 2.82 & 0.018 & 37.00 & 3.4400 & 0.0584 & 0.0941 \\
	A43  & 0.6892 & 7994 & 2.20 & 0.2590 & 3.8787 & 29.35 & 2.82 & 0.020 & 30.00 & 3.3983 & 0.0677 & 0.0956 \\
	A44  & 0.7629 & 7689 & 2.14 & 0.2583 & 3.8730 & 24.76 & 2.80 & 0.023 & 31.00 & 3.3888 & 0.0671 & 0.0958 \\
	A45  & 0.6790 & 8026 & 2.21 & 0.2602 & 3.8797 & 29.89 & 2.83 & 0.020 & 42.00 & 3.3985 & 0.0679 & 0.0951 \\
	A46  & 0.7369 & 7791 & 2.16 & 0.2589 & 3.8751 & 26.21 & 2.81 & 0.022 & 30.00 & 3.3910 & 0.0672 & 0.0907 \\
	A47  & 0.7369 & 7791 & 2.16 & 0.2589 & 3.8751 & 26.21 & 2.81 & 0.022 & 32.00 & 3.3910 & 0.0672 & 0.0980 \\
	A48  & 0.7369 & 7791 & 2.16 & 0.2589 & 3.8751 & 26.21 & 2.81 & 0.022 & 31.00 & 3.3910 & 0.0672 & 0.0870 \\
	A49  & 0.7125 & 7892 & 2.18 & 0.2592 & 3.8769 & 27.74 & 2.82 & 0.021 & 30.00 & 3.3946 & 0.0673 & 0.0958 \\
	A50  & 0.7125 & 7892 & 2.18 & 0.2592 & 3.8769 & 27.74 & 2.82 & 0.021 & 31.00 & 3.3946 & 0.0673 & 0.0835 \\
	A51  & 0.7125 & 7892 & 2.18 & 0.2592 & 3.8769 & 27.74 & 2.82 & 0.021 & 32.00 & 3.3946 & 0.0673 & 0.0856 \\
	A52  & 0.6889 & 7995 & 2.20 & 0.2592 & 3.8790 & 29.34 & 2.82 & 0.020 & 31.00 & 3.3965 & 0.0675 & 0.0962 \\
	A53  & 0.7475 & 7760 & 2.15 & 0.2578 & 3.8748 & 25.70 & 2.80 & 0.022 & 30.00 & 3.3870 & 0.0667 & 0.0900 \\
	A54  & 0.7475 & 7760 & 2.15 & 0.2578 & 3.8748 & 25.70 & 2.80 & 0.022 & 37.00 & 3.3870 & 0.0667 & 0.0900 \\
	A55  & 0.6986 & 7965 & 2.19 & 0.2582 & 3.8787 & 28.80 & 2.82 & 0.020 & 32.00 & 3.3929 & 0.0669 & 0.0988 \\
	A56  & 0.7331 & 7832 & 2.16 & 0.2573 & 3.8763 & 26.70 & 2.81 & 0.021 & 37.00 & 3.3869 & 0.0667 & 0.0865 \\
	A57  & 0.7331 & 7832 & 2.16 & 0.2573 & 3.8763 & 26.70 & 2.81 & 0.021 & 38.00 & 3.3869 & 0.0667 & 0.0820 \\
	A58  & 0.7987 & 7755 & 2.15 & 0.3017 & 3.8649 & 26.22 & 2.84 & 0.024 & 30.00 & 3.4404 & 0.0699 & 0.0973 \\
	A59  & 0.7987 & 7755 & 2.15 & 0.3017 & 3.8649 & 26.22 & 2.84 & 0.024 & 31.00 & 3.4404 & 0.0699 & 0.0969 \\
	A60  & 0.7860 & 7903 & 2.15 & 0.2923 & 3.8642 & 28.32 & 2.84 & 0.020 & 30.00 & 3.4530 & 0.0685 & 0.0852 \\
	A61  & 0.7860 & 7903 & 2.15 & 0.2923 & 3.8642 & 28.32 & 2.84 & 0.020 & 31.00 & 3.4530 & 0.0685 & 0.0941 \\
	A62  & 0.7978 & 7873 & 2.14 & 0.2917 & 3.8637 & 27.80 & 2.83 & 0.020 & 30.00 & 3.4502 & 0.0684 & 0.0960 \\
	A63  & 0.7978 & 7873 & 2.14 & 0.2917 & 3.8637 & 27.80 & 2.83 & 0.020 & 33.00 & 3.4502 & 0.0684 & 0.0878 \\
	A64  & 0.7978 & 7873 & 2.14 & 0.2917 & 3.8637 & 27.80 & 2.83 & 0.020 & 34.00 & 3.4502 & 0.0684 & 0.0945 \\
	A65  & 0.7978 & 7873 & 2.14 & 0.2917 & 3.8637 & 27.80 & 2.83 & 0.020 & 31.00 & 3.4502 & 0.0684 & 0.0925 \\
	A66  & 0.7978 & 7873 & 2.14 & 0.2917 & 3.8637 & 27.80 & 2.83 & 0.020 & 32.00 & 3.4502 & 0.0684 & 0.0851 \\
	A67  & 0.9066 & 7605 & 2.05 & 0.2857 & 3.8633 & 23.21 & 2.78 & 0.020 & 32.00 & 3.3981 & 0.0667 & 0.0958 \\
	A68  & 0.9066 & 7605 & 2.05 & 0.2857 & 3.8633 & 23.21 & 2.78 & 0.020 & 34.00 & 3.3981 & 0.0667 & 0.0988 \\
	A69  & 0.9066 & 7605 & 2.05 & 0.2857 & 3.8633 & 23.21 & 2.78 & 0.020 & 33.00 & 3.3981 & 0.0667 & 0.0955 \\
	A70  & 0.7255 & 8127 & 2.20 & 0.2932 & 3.8687 & 32.08 & 2.86 & 0.019 & 42.00 & 3.4610 & 0.0695 & 0.0805 \\
	A71  & 0.7255 & 8127 & 2.20 & 0.2932 & 3.8687 & 32.08 & 2.86 & 0.019 & 41.00 & 3.4610 & 0.0695 & 0.0933 \\
	A72  & 0.8635 & 7747 & 2.08 & 0.2808 & 3.8612 & 25.48 & 2.80 & 0.019 & 30.00 & 3.4321 & 0.0666 & 0.0827 \\
	A73  & 0.8290 & 7879 & 2.10 & 0.2792 & 3.8638 & 27.35 & 2.81 & 0.018 & 30.00 & 3.4333 & 0.0667 & 0.0870 \\
	A74  & 0.8290 & 7879 & 2.10 & 0.2792 & 3.8638 & 27.35 & 2.81 & 0.018 & 31.00 & 3.4333 & 0.0667 & 0.0965 \\
	A75  & 0.8764 & 7716 & 2.07 & 0.2800 & 3.8607 & 24.98 & 2.80 & 0.019 & 30.00 & 3.4285 & 0.0664 & 0.0649 \\
	A76  & 0.8413 & 7847 & 2.09 & 0.2783 & 3.8633 & 26.82 & 2.80 & 0.018 & 30.00 & 3.4303 & 0.0666 & 0.0737 \\
	A77  & 0.8846 & 7701 & 2.06 & 0.2825 & 3.8650 & 24.43 & 2.78 & 0.019 & 40.00 & 3.3988 & 0.0667 & 0.0999 \\
	A78  & 0.8542 & 7816 & 2.08 & 0.2775 & 3.8626 & 26.31 & 2.80 & 0.018 & 30.00 & 3.4283 & 0.0665 & 0.0731 \\
	A79  & 0.8979 & 7670 & 2.05 & 0.2816 & 3.8646 & 23.94 & 2.77 & 0.019 & 39.00 & 3.3947 & 0.0666 & 0.0973 \\
	A80  & 0.8979 & 7670 & 2.05 & 0.2816 & 3.8646 & 23.94 & 2.77 & 0.019 & 36.00 & 3.3947 & 0.0666 & 0.0997 \\
	A81  & 0.8979 & 7670 & 2.05 & 0.2816 & 3.8646 & 23.94 & 2.77 & 0.019 & 37.00 & 3.3947 & 0.0666 & 0.0903 \\
	A82  & 0.8979 & 7670 & 2.05 & 0.2816 & 3.8646 & 23.94 & 2.77 & 0.019 & 38.00 & 3.3947 & 0.0666 & 0.0872 \\
	A83  & 0.8887 & 7738 & 2.05 & 0.2781 & 3.8657 & 24.74 & 2.77 & 0.018 & 34.00 & 3.3930 & 0.0664 & 0.0996 \\
	A84  & 0.8887 & 7738 & 2.05 & 0.2781 & 3.8657 & 24.74 & 2.77 & 0.018 & 35.00 & 3.3930 & 0.0664 & 0.0831 \\
\end{longtable}


\bibliography{Ref}{}
\bibliographystyle{aasjournal}



\end{document}